\documentclass{emulateapj}

\usepackage{amsmath,amssymb,verbatim} 
\usepackage{color} 
\usepackage{natbib}

\shorttitle{Accretion Flows}
\shortauthors{}

\begin{document}
\title{Rotating Accretion Flows: From Infinity to the Black Hole}


\author{Jason Li$^1$, Jeremiah Ostriker$^1$, \& Rashid Sunyaev$^{2,3,4}$}
\affil{$^1$Department of Astrophysical Sciences, Peyton Hall,
Princeton University, Princeton, NJ 08544, USA \\ $^2$ Institute for Advanced Study, Einstein Drive, Princeton, New Jersey 08540, USA \\ $^3$ Max Planck
Institut f\"{u}r Astrophysik, Karl-Schwarzschild-Str. 1 85741, Garching,
Germany \\ $^4$ Space Research Institute, Russian Academy of Sciences, Profsoyuznaya 84/32, 117997
Moscow, Russia \\  } \email{jgli@astro.princeton.edu}

\begin{abstract}

Accretion onto a supermassive black hole of a rotating inflow is a
particularly difficult problem to study because of the wide range of
length scales involved.  There have been broadly utilized analytic and
numerical treatments of the global properties of accretion flows, but
detailed numerical simulations are required to address certain
critical aspects.  We use the ZEUS code to run hydrodynamical
simulations of rotating, axisymmetric accretion flows with
Bremsstrahlung cooling, considering solutions for which the
centrifugal balance radius significantly exceeds the Schwarzschild
radius, with and without viscous angular momentum transport.
Infalling gas is followed from well beyond the Bondi radius down to
the vicinity of the black hole.  We produce a continuum of solutions
with respect to the single parameter $\dot{M}_{\rm B}/\dot{M}_{\rm
Edd}$, and there is a sharp transition between two general classes of
solutions at an Eddington ratio of $\dot{M}_{\rm B}/\dot{M}_{\rm
Edd}\sim{\rm few}\times 10^{-2}$.  Our high inflow solutions are very
similar to the standard Shakura \& Sunyaev (1973) results. But our low
inflow results are to zeroth order the stationary Papaloizou and
Pringle (1984) solution, which has no accretion.  To next order in the
small, assumed viscosity they show circulation, with disk and conical
wind outflows almost balancing inflow.  These solutions are
characterized by hot, vertically extended disks, and net accretion
proceeds at an extremely low rate, only of order $\alpha$ times the
inflow rate.
Our simulations have converged
with respect to spatial resolution and temporal duration, and they do
not depend strongly on our choice of boundary conditions.

\end{abstract}




\section{Introduction}
Supermassive Black Hole accretion is affected by the properties of the
accretion flow on a wide range of length scales.  Gravity and gas
dynamics play an important role from the gravitational radius out to
much larger distances where the gravitational potential energy of the
gas is comparable to its kinetic energy.  The properties of the
infalling gas at even larger radii, where gravity is weak can also in
principle affect the global properties of the accretion flow, an
important point that is sometimes overlooked in studies of accretion
physics.

In the unmagnetized spherically symmetric case with no angular
momentum and adiabatic equation of state, the global accretion flow is
given by the Bondi solution \citep{bondi52}.  The Bondi radius $R_{\rm
B}\equiv GM_{\rm BH}/c_{s,\infty}^2$ sets the scale at which gravity
becomes important, and the mass accretion rate is given by
$\dot{M}_{\rm B}\equiv \Lambda4\pi r_{\rm B}^2\rho_\infty
c_{s,\infty}$.  $M_{\rm BH}$ here is the black hole mass,
$\rho_\infty$ is the mass density of gas at infinity, $c_{s,\infty}$
is the gas sound speed at infinity, and
$\Lambda\equiv[2/(5-3\gamma)]^{(5-3\gamma)/[2(\gamma-1)]}/4$
\citep{frank02}, where $\gamma$ is the adiabatic index.  The
introduction of angular momentum fundamentally modifies this picture,
however, and for sufficiently high angular momentum the gas can no
longer accrete onto the central black hole.  For gas with specific
angular momentum $j$, the centrifugal radius $R_{\rm c}\equiv
j^2/GM_{\rm BH}$ defines the radius interior to which the gas has
difficulty penetrating.  Accretion in this case requires some form of
angular momentum transport.  Turbulent stresses associated with the
nonlinear development of the MagnetoRotational Instability (MRI) are
thought to be capable of transporting angular momentum
\citep{velikhov59,chandrasekhar60,balbus91}.  The angular momentum transport
properties of the MRI are often parameterized with an effective shear
viscosity $\nu\equiv \alpha c_s H_p$, where numerical simulations
suggest $\alpha$ is typically of order $0.01$ \citep{stone96,hawley11,mckinney12}.
Observations actually suggest a vertically and azimuthally averaged
$\alpha \sim 0.1-0.4$ \citep{king07}, a discrepancy that to date has
not yet been fully addressed.

There are a number of general classes of analytic solutions for
rotating flows with viscous angular momentum transport.  The standard
solution in the limit of a geometrically thin and optically thick
viscous disk was given by Shakura \& Sunyaev (1973; see also
Lynden-Bell \& Pringle 1974, Novikov \& Thorne 1973), commonly
referred to as an $\alpha$-disk.  Thin disks typically neglect global
heat transport and only consider effects that are first order in the
disk thickness $H/R$, where $H$ is the disk thickness at cylindrical
radius $R$.  The next order of approximation is slim disks, which
account for terms in the equations of motion that are second order in
the disk thickness $H/R$ (Abramowicz et al. 1988; see also Abramowicz
et al. 1986, Kato et al. 1988, Chen \& Taam 1993, Narayan \& Popham
1993, Katz 1977, Begelman 1978, Begelman \& Meier 1982, Eggum et
al. 1988).  Slim disks allow for non-negligible radial velocity,
horizontal pressure gradients, and advective heat transport, and they
have been used to model geometrically thick, hot disks accreting at
super-Eddington rates.  Advection Dominated Accretion Flows
\citep[ADAFs,][]{narayan94,narayan95a,
narayan95b,abramowicz95,ichimaru77,rees82,quataert99a,narayan11} describe
radiatively inefficient sub-Eddington flows.  The accreting low density
gas is hot and geometrically extended in the polar direction, and the thermal energy is advected onto the central
object.  The closely related Adiabatic Inflow Outflow Solutions
\citep[ADIOS,][]{blandford99,begelman11} can accrete at much lower
rates by allowing for outflows that can carry off energy and angular
momentum \citep[see also][]{jiao11}.  At present it is unclear which,
if any, of these solutions real supermassive black hole accretion
flows will pick.  Certainly for hot disks with $H\sim R$, the standard
approximations in thin disk accretion theory break down
\citep{pringle81}, and we should consider terms of all order in $H/R$.
Further study of the detailed properties of geometrically
thick disks is required.

Numerical simulations have provided further insight on accretion physics, 
and Convection Dominated
Accretion Flows \citep[CDAFs;][]{igumenshchev99,igumenshchev00,igumenshchev00b,igumenshchev03,stone99,narayan00,quataertgruzinov00}
, which operate in the weak viscosity regime for
flows with no cooling, have been proposed to explain one class of simulations.  Recent work by \cite{yuan12b} tends to favor ADIOS-type models over CDAFs, however.  Other authors have studied the geometry of two
dimensional magnetohydrodynamical flows accreting from torus-like
initial reservoirs of gas
(\cite{stone01,hawley01,machida01,igumenshchev02,hawley02}; see also
\cite{narayan12} for 3D simulations with general relativity).  Many previous works neglect the radiative properties of the flow, however, which may play
an important role in determining the geometry of the flow \citep{dibi12,fragile12}.  For
accretion flows with radiative cooling, the properties of the
infalling gas at infinity determine the strength of the radiative
cooling and can affect the global properties of the flow (Yuan et
al. 2000).  Several authors mimic cooling by varying the adiabatic
index of the gas \citep{moscibrodzka08,moscibrodzka09,janiuk09} or
include radiative transfer (\citealt{ohsuga09,ohsuga11}; see also \citealt[][and references
therein]{zanotti11}), but accretion in these works is aided by at least
some inflowing gas that has low angular momentum or starting with gas that is bound to the black hole.  Another important
consequence of radiation from gas near the black hole is that it can
drastically transform the outer parts of the accretion flow via
heating \citep{park01,park07}.

It is well known that observed black holes can accrete at rates that
are only a small fraction of the Bondi accretion rate.  For example,
Sgr ${\rm A}^*$ appears to be accreting at a rate of
$10^{-3}-10^{-2}\dot{M}_{\rm B}$ \citep{yuan03,quataert04}.  A number
of authors have studied the transition between the cooling flow at
large radii and the accretion flow near and below the Bondi radius
\citep{quataert99b,quataertnarayan00}, and the implications for
accretion rates.  Various energy transport mechanisms have been
suggested that may be able to reduce accretion rates well below the
Bondi rate
\citep{ciotti01,quataert04,sharma08,shcherbakov10,igumenshchev06}.
Other authors have invoked Radiatively Inefficient Accretion Flows to
explain the low luminosity of observed accreting systems \citep[][and
references therein]{narayan08}.
We consider here the alternate view that it is simply the angular
momentum of the gas that slows accretion.  There has been some
analytic and semi-analytic work exploring this idea, including 
angular momentum transport \citep{xu97,blandford99}, and
both angular momentum transport and cooling
\citep[e.g.,][]{chakrabarti96,park09,inogamov10}, as well as numerical
work without cooling \citep[e.g.,][]{stone99,proga03a,proga03b,yuan12a,yuan12b} and
with cooling \citep{yuan10}.

This paper will treat the accretion problem in the domain where the
angular momentum is not sufficient to prevent inflow at the Bondi rate
but ample to inhibit accretion without the operation of some viscous
process.  That is, we will work exclusively in the domain
\begin{equation}
R_s \ll R_{\rm c} \ll R_{\rm B},
\label{eq:rcent}
\end{equation}
where $R_s\equiv 2GM_{\rm BH}/c^2$ is the Schwarzschild radius.  Since the
ratio of the bounding radii, $R_{\rm B}/R_s \sim
10^{8}/T_4$, this choice is well-defined.
Moreover, many elliptical galaxies, including M87, 
have stellar populations with small average spin that is insufficient to create Keplerian disks 
near the Bondi radius \citep[see e.g.,][]{inogamov10}.
We run hydrodynamical simulations to follow infalling, radiating gas that
starts from well beyond the Bondi radius with
$Be>c_{s,\infty}^2/(\gamma-1)$ (see Equation \ref{eq:bernoulli}) down
to the vicinity of the black hole.  
We include shear viscosity to capture
qualitatively the angular momentum transport properties of the MRI,
and so accretion can proceed at varying rates, depending on the flow
parameters.  Since the gravitational and centrifugal forces do not
have the same scaling with radial distance in our setup, we do not
expect our solutions to be self-similar as in standard ADAF
\citep{narayan94} and CDAF models\citep{quataertgruzinov00,narayan00}.
The thermal Bremsstrahlung cooling can also play an essential role in
the energetics of the flow and determine if gas is bound to the black
hole or not.  
We provide accretion flow solutions in both the very sub-Eddington,
nearly adiabatic domain, and also solutions approaching the
super-Eddington domain.  Section \ref{sec:method} provides the
mathematical framework framework for our problem and describes our
numerical method.  Section \ref{sec:results} gives our main results,
and the broader implications of our results are discussed in sections
\ref{sec:discussion} and \ref{sec:conclusions}.  The Appendix provides
details on the convergence properties of our solutions.  We will
designate the low inflow solutions that we find as ``RRIOS'' -
Radiating Rotating Inflow-Outflow Solutions.

\section{Method}\label{sec:method}

\subsection{Equations}\label{sec:eqs}
We use ZEUS-2D v2.0 (Stone \& Norman 1992) to solve the equations of
hydrodynamics,
\begin{align}
\left(\frac{\partial }{\partial t}+ \boldsymbol{v\cdot\nabla} \right)\rho+ \rho \boldsymbol{\nabla\cdot v} =0, \\
\rho\left(\frac{\partial }{\partial t}+ \boldsymbol{v\cdot\nabla} \right)\boldsymbol{v}=-\boldsymbol{\nabla}p - \rho \boldsymbol{\nabla}\Phi, \\
\rho\left(\frac{\partial }{\partial t}+ \boldsymbol{v\cdot\nabla} \right)\frac{e}{\rho}=-p\boldsymbol{\nabla\cdot v},
\end{align}
where $\rho$ is mass density, $\boldsymbol{v}$ is flow velocity, $e$
is energy density, gas pressure $p\equiv (\gamma-1)e$, where $\gamma$
is the adiabatic index, and the gravitational potential $\Phi\equiv
-GM_{\rm BH}/r$.  We run global hydrodynamical simulations of
axisymmetric rotating accretion flows in spherical polar coordinates.
The simulations include thermal Bremsstrahlung cooling, shear
viscosity to capture drag due to turbulent stresses, and modified
boundary conditions to allow for disk and wind outflows.  The source
terms for momentum and energy in the ZEUS hydrocode are modified by
the addition of the terms within square brackets below:
\begin{equation}
  \label{eq:NavierStokesNumerical}
\rho\left(\frac{\partial \boldsymbol{v}}{\partial t}\right)_{\rm sources}
= -\boldsymbol{\nabla} p -\rho \boldsymbol{\nabla} \Phi  -\nabla \cdot \boldsymbol{Q}+
\left[\boldsymbol{\nabla\cdot\sigma'}\right]\,,
\end{equation}
and
\begin{equation}
  \label{eq:energy}
\begin{split}
\left(\frac{\partial e}{\partial t}\right)_{\rm sources}= &-p
\boldsymbol{\nabla\cdot v} -\boldsymbol{Q\cdot\nabla v} + \\
& \left[\boldsymbol{\sigma'\cdot\nabla v} - \dot{e}_{\rm Brem} 
\right]\,,
\end{split}
\end{equation}
in which $\boldsymbol{\sigma'}$ is the viscous shear tensor and
$\boldsymbol{q}$ is the heat flux due to thermal conduction of
electrons.
ZEUS' standard tensor artificial viscosity $\boldsymbol{Q}$ is applied
with shocks spread over $\approx 2$ zones, and we use a Foward Time
Centered Space differencing scheme for our modified source terms.  We
neglect the non-azimuthal components of the viscous shear tensor
$\boldsymbol{\sigma'}$, as it is believed that the poloidal shear is
subdominant to the azimuthal shear in the non-linear regime of the MRI
\citep[see][]{stone99}.  The azimuthal components of the viscous shear
tensor are
\begin{equation}
\boldsymbol{\sigma'}_{r\phi}=\nu \rho r \frac{\partial}{\partial r}\left(\frac{v_{\phi}}{r}\right),
\end{equation}
\begin{equation}
\boldsymbol{\sigma'}_{\theta\phi}=\frac{\nu \rho \sin \theta}{r} \frac{\partial}{\partial \theta}\left(\frac{v_{\phi}}{\sin \theta}\right),
\end{equation}
where $\nu$ is the kinematic viscosity.
The Bremsstrahlung
cooling term (Svensson 1982, Ball et al. 2001)
\begin{equation}
\dot{e}_{\rm Brem}\equiv \alpha_f
r_e^2m_ec^3n^2(32/3)(2/\pi)^{1/2}(\frac{k_BT}{m_ec^2})^{1/2},
\label{eq:brem}
\end{equation}
where $\alpha_f$ is the fine structure constant and $r_e$ is the
classical electron radius.  We use this abbreviated form for the
Bremsstrahlung cooling to capture the qualitative effects of thermal
emission; in fact in high temperature regions the electron component
approaches relativistic speeds and Bremsstrahlung radiation increases
substantially above equation (\ref{eq:brem}).  The cooling is limited
by a floor in the temperature, equal to the gas temperature at the
outer boundary, in order to maintain stability.  The advective
transport terms in ZEUS are unmodified.
\subsection{Setup \& Flow Properties}
In our simulations gas flows in through the outer boundary, located at
$R_{\rm out}=10R_B$, at the Bondi rate, $\dot{M}_{\rm B}$.  We modify
the outer boundary condition locally to an outflow boundary condition
in whichever grid zones at the outer boundary that the flow is moving
radially outward, however.  We note that several other authors have
studied accretion flows with outer boundary near the Bondi radius
\citep[e.g.,][]{pen03,pang11}.  Our inner boundary, located at $R_{\rm
in}=10^{-3} R_B$, is a standard no torque outflow boundary condition.
$R_{\rm in}$ is located at $91R_s$, and we have checked the
convergence of our results with respect to the location of the inner
boundary (see Appendix).  The flow is initialized throughout with
Bondi profile for density, energy density, and radial velocity.  Our
standard simulations are run with $T_{\infty}=2\times 10^7$ K.  This
is close to the temperature of gas heated radiatively by a typical
quasar output spectrum \citep{sazonov05}.  The simulations are run
with adiabatic index $\gamma=1.5$, except where otherwise noted, and
we verify explicitly that this assumption does not affect our main
results (cf Figure \ref{fig:gamma}).  The real gas will have
$\gamma=5/3$ above $R\sim 300-500 R_s$.  At lower radii, in the
non-collisional case, electrons become relativistic while protons are
nonrelativistic, leading to lower effective $\gamma$.
Our choice for $\gamma$ is made in
order to allow us to well separate the sonic and centrifugal radii for
supersonic infall.  The density at infinity, $\rho_\infty$, is a
parameter that determines the dimensionless quantity $\dot{M}_{\rm
B}/\dot{M}_{\rm Edd}$, where $\dot{M}_{\rm Edd}\equiv 4\pi G M_{\rm
BH}m_p / \epsilon \sigma_T c$ is the Eddington luminosity and we
assume an efficiency $\epsilon=0.1$.  This dimensionless mass
accretion rate parameter scales as $\dot{M}_{\rm B}/\dot{M}_{\rm
Edd}\propto M_{\rm BH}\rho_\infty /c_{s,\infty}^3$.  The dimensionless
cooling fraction $\dot{e}_{\rm Brem} \Delta t_{\rm char}/nkT \propto
M_{\rm BH}\rho_\infty /c_{s,\infty}^4$, where we have taken
characteristic time $\Delta t_{\rm char}=M_{\rm BH}/c_{s,\infty}^4$.
The black hole mass and gas density at infinity appear together in the
above two quantities, and the free parameter in the accretion flow is
actually the product $M_{\rm BH}\rho_{\infty}$ (see Chang \& Ostriker
1985).  This parameter effectively determines the strength of the
Bremsstrahlung cooling, with $\dot{M}_{\rm B}/\dot{M}_{\rm Edd} \ll 1$
corresponding to no cooling and $\dot{M}_{\rm B}/\dot{M}_{\rm Edd} \gg
1$ corresponding to strong cooling.  For convenience we pick $M_{\rm
BH}=2\times10^8 M_{\odot}$ and allow $\rho_\infty$ to vary freely.  At
$\rho_\infty=2.3\times10^{-21} g\, cm^{-3}$, the Eddington ratio
$\dot{M}_{\rm B}/\dot{M}_{\rm Edd}=0.1$.

Gas is initialized with constant specific angular momentum throughout,
except near the rotation axis, where it tapers to zero.  The inflowing
gas at the outer boundary has the same angular momentum, without
tapering, ensuring that after initial transients, the gas has uniform
angular momentum throughout.  For adiabatic inviscid
flows, the Bernoulli constant
\begin{equation}
Be\equiv  \frac12\left(v_r^2+v_\theta^2\right)+\frac12\frac{j^2}{\varpi^2}-\frac{GM}{r}+\frac{c_s^2}{\gamma-1},
\label{eq:bernoulli}
\end{equation}
where $\varpi$ is cylindrical radius, is a conserved quantity (see
e.g., Narayan \& Yi 1994).  It immediately follows that there is a
centrifugal radius interior to which the gas cannot penetrate if gas
starts from rest at infinity with $Be=\gamma
c_{s,\infty}^2/(\gamma-1)$.  The specific angular momentum is another
free parameter in our simulations, but we pick
$j_\infty=\sqrt{0.02}R_{\rm B}c_{s,\infty}$ so that the centrifugal
radius $R_{\rm c}=0.02 R_{\rm B}$ is well resolved and lies at a
radius $20$ times that of the inner boundary.  Hence we are computing
within the domain given by equation (\ref{eq:rcent}).  The limit of
weak angular momentum with $R_{\rm c}\ll R_{\rm in}$ approaches the
Bondi solution.  For our choice of specific angular momentum, the
introduction of shear viscosity and angular momentum transport allows
for gas to migrate from the centrifugal radius down to the inner
boundary and to accrete.  The accreting gas carries only a small
fraction ($<20\%$) of the initial angular momentum $j_\infty$ through
the inner boundary.  In our simulations we set the kinematic viscosity
$\nu$ to a constant nonzero value $\nu=10^{-3}c_{s,\infty}R_{\rm B}$.
The effective $\alpha$ coefficient, $\alpha\equiv \nu/c_sH_p$, where
$c_s$ is the local sound speed and $H_p$ the local pressure scale
height (see Section \ref{sec:radcool}), is then spatially variable,
but is typically of order $\sim 0.01$ near the centrifugal barrier
where the density peaks.  \cite{stone99} found that the properties of
their non-radiative hydrodynamic flow do not depend strongly on the
assumed form for kinematic viscosity, and \cite{stone01} found weakly
magnetized flows to be qualitatively similar to the viscous
hydrodynamic ones.  Incidentally, the limit of large specific angular
momentum with $R_{\rm c} \gtrsim R_{\rm B}$ should approach that of a
giant viscous disk with large radial extent, but we have not explored
this regime in any detail.  The Bremsstrahlung cooling also modifies
the flow profile, and for cold disks the inflow can be highly
supersonic.  In these cold solutions the sonic radius is typically
intermediate between but still well separated from the centrifugal and
Bondi radii.

\subsection{Numerics}
\label{sec:numerics}
Our standard simulations are run with $21$ equally spaced grid zones in polar
angle $\theta$ varying from $\theta=0$ to $\theta=\pi/2$, enforcing
symmetry across the equatorial plane.  In the radial direction we use
a non-uniform logarithmic grid with $32$ zones per decade ranging from
$R_{\rm in}=10^{-3}R_{\rm B}$ to $R_{\rm out}=10R_{\rm B}$,
with $128$ total grid zones.  As a code check, we ran the spherically
symmetric test problem with no cooling.  Our code is able to maintain
the Bondi profile for mass density and radial velocity to within $5\%$
for various values of the adiabatic index $\gamma$.  The internal
energy density is less well conserved in the very inner portions of
the flow ($\approx 10$ innermost cells) due to artificial viscous
heating, but is still within a factor of $\approx 2$ of the Bondi
solution.  The simulations are run until all time-averaged gross
quantifiers of the flow structure approach quasi-steady values.  All
diagnostics of the flow are averaged over many tens of Bondi times
$t_{\rm B}\equiv R_{\rm B}/c_{s,\infty}$, typically between
times $t=40\, t_{\rm B}$ and $t=90\,t_{\rm B}$ (see Appendix).  Flow patterns
are illustrated at time $t=90\,t_{\rm B}$ except where otherwise
noted.  The code is mass conservative, so mass flowing inwards from
large radii must accumulate on the grid or outflow either through the
inner boundary or back out to large radii via the combination of polar
and equatorial outflow.  

We also run a number of tests to check the proper implementation of
the additional physics we have added to ZEUS.  We verified the proper
thermal cooling of a stationary constant density gas, as well as the
diffusion of thermal energy when a temperature gradient exists.
Finally, we have reproduced qualitatively the viscous evolution of an
initial ring of matter \citep{pringle81}.  The ring spreads in radius
as matter moves inwards, but the bulk of the angular momentum is
transferred outwards.  Kinetic energy lost to viscous dissipation is
added to the internal energy of the gas, with the viscous substep of
the code conserving energy to within $10$\%.

\section{Results}\label{sec:results}

\subsection{Adiabatic Inviscid Flows}

For adiabatic flows with sufficiently high angular momentum that the
centrifugal barrier is located far from the black hole, equation
(\ref{eq:bernoulli}) immediately tells us that there can be no
accretion onto the black hole.  Any inflowing high angular momentum
material must either accumulate beyond the centrifugal radius or be
deflected back outwards and ejected in outflows (\citealt[see][and
references therein]{proga03a}; see also
\citealt{hawley84a,hawley84b,hawley86}).

In the adiabatic regime with constant angular momentum $j$ throughout,
there is a stationary ($v_r=v_\theta=0$) settling solution satisfying the
momentum equation \citep[see][]{papaloizou84,fishbone76}.  Starting from the
momentum equation,
\begin{equation}
-\frac{1}{\rho}\boldsymbol{\nabla}p - \boldsymbol{\nabla}\Phi
 +\frac{j^2}{\varpi^3} \boldsymbol{\hat{\varpi}}=0,
\end{equation}
the adiabatic relation $p\propto \rho^{\gamma}$ gives
\begin{equation}
\boldsymbol{\nabla} \left(\frac{c_s^2}{\gamma-1} + \Phi
 +\frac{j^2}{2\varpi^2} \right) =0.
\end{equation}
This yields an equation for the sound speed,
\begin{equation}
\frac{c_s^2}{c_{s,\infty}^2}=1+(\gamma-1)\frac{GM}{c_{s,\infty}^2r}-\frac{\gamma-1}{2}\frac{j^2}{c_{s,\infty}^2\varpi^2},
\label{eq:cs}
\end{equation}
valid where the right hand side is positive,
as well as the density profile,
\begin{equation}
\frac{\rho}{\rho_{\infty}}=[1+(\gamma-1)\frac{GM}{c_{s,\infty}^2r}-\frac{\gamma-1}{2}\frac{j^2}{c_{s,\infty}^2\varpi^2}]^{1/(\gamma-1)}.
\label{eq:dens}
\end{equation}
The density is zero near the polar axis in the region where the right
hand side of equation (\ref{eq:cs}) is negative.  This solution is
actually dynamically unstable to non-axisymmetric perturbations
\citep{papaloizou84}, which we do not allow for in our axisymmetric
code.  Neglecting these instabilities is likely a reasonable
assumption, however, as we have angular momentum gradients that can
stabilize the flow \citep{papaloizou87}.

Our multidimensional adiabatic simulations closely resemble this
stationary settling solution, with sound speed profile matching
equation (\ref{eq:cs}) to within a few percent exterior to the
centrifugal barrier.  Figure~1a shows the structure of the flow within
the Bondi radius at Eddington ratio $\dot{M}_{\rm B}/\dot{M}_{\rm
Edd}=10^{-4}$.  The red contours are logarithmic contours of density,
given in cgs units, and we overlay black contours of logarithmic
density for the adiabatic settling solution.  The solid black line
shows the zero density surface in the stationary solution.  Blue
arrows indicate vectors of mass flux in the $R-z$ plane, with length
proportional to the magnitude of the mass flux vector.  Subsequent
diagrams of flow patterns showing logarithmic density contours and
mass flux vectors are illustrated in a similar fashion.  The pressure
associated with the accumulation of matter in an extended torus
impedes the inflow of matter, as in the stationary solution.  Indeed,
the mass inflow rate at the Bondi radius is reduced by more than one order of
magnitude to $\sim 0.06\dot{M}_{\rm B}$ at $t=45 t_{\rm
B}$.  There is evidence of time-dependent circulation,
however, partially due to our choice of initial conditions, and the
flow is not purely stationary with zero inflow rate at the Bondi
radius.  Matter flows inwards at intermediate latitudes and turns around at radii exterior to the centrifugal barrier, flowing back outwards in the polar and
equatorial regions.  Figure~1b shows the flow structure interior to
$0.1R_{\rm B}$, with centrifugal radius located at $R_{\rm
c}=0.02 R_{\rm B}$.  There is again evidence of time-dependent
circulation, but the largest differences in the flow relative to the
stationary solution occur near the polar axis where the stationary
solution has zero mass density.  There are very small inflows in the
polar region due to angular momentum transport associated with
numerical viscosity, but the accretion rate is extremely low,
$\dot{M}_{\rm accretion}< 10^{-3}\dot{M}_{\rm B}$, i.e. any
accretion is due to numerical error.  In the absence of physical
angular momentum transport the gas in the centrifugal barrier is
stationary in the $r$ and $\theta$ directions, cannot accrete, and its
pressure prevents new gas from falling down to small radii.  Our
results indicate that in the limit of very long and accurate
integrations the solution would approach the stationary solution given
in equations (\ref{eq:cs}) and (\ref{eq:dens}), since there is no
engine to drive circulation for adiabatic flows.

\begin{figure*}[htp]
\centering
\includegraphics[width=.87\textwidth]{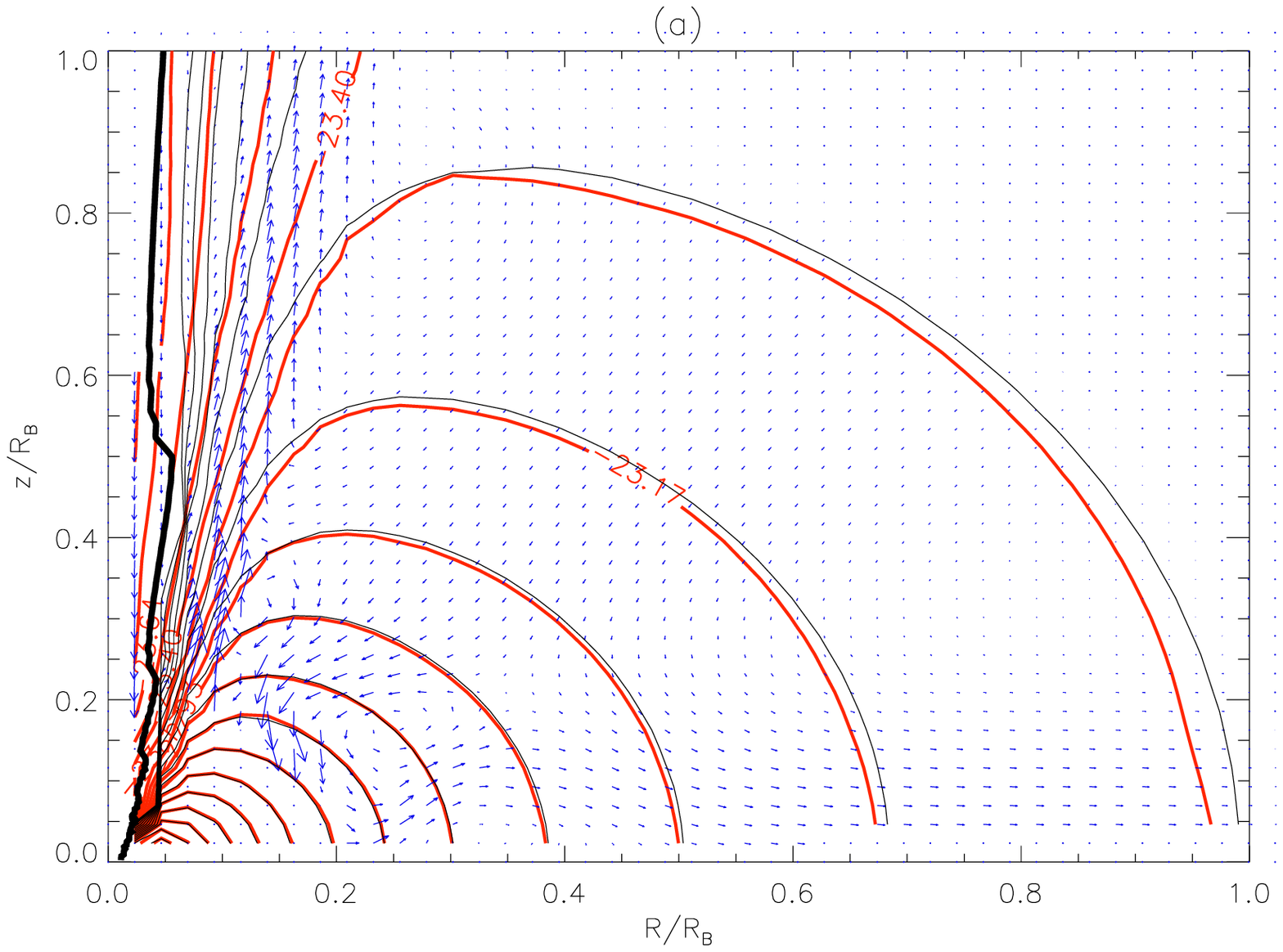}\\
\includegraphics[width=.87\textwidth]{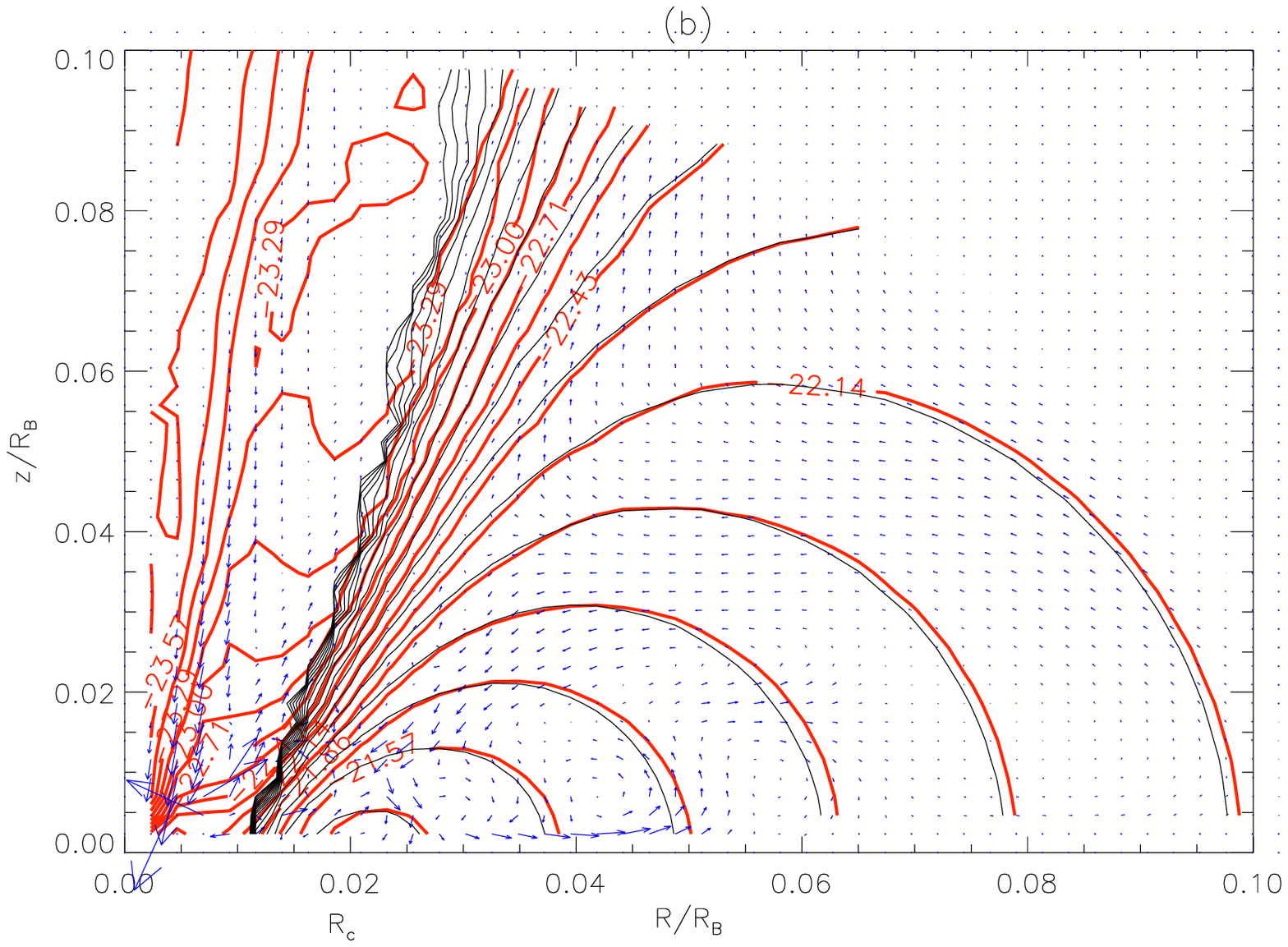}
\caption{(a) The structure of the inviscid flow with no cooling,
interior to $R_{\rm B}$ at $t=45 t_{\rm B}$.  Red
contours are logarithmic contours of density, and blue arrows indicate
mass flux in the $R-z$ plane.  The flow approaches a stationary
settling solution (black contours, cf equation (\ref{eq:dens})) in
which the inflow rate through the Bondi radius is reduced by an order
of magnitude below the Bondi rate.  The solid black line shows the zero density surface in the stationary solution.  (b) Flow structure interior to
$5R_{\rm c}=0.1R_{\rm B}$.  The bulk of the gas in the disk
cannot accrete and the accretion through the polar funnel is
negligible and due to numerical error.}\label{fig:inviscid}
\end{figure*}

\subsection{Hot and Cold Disks}\label{sec:hotcold}

We find two general states for rotating axisymmetric accretion flows
with radiative cooling and viscous angular momentum transport: hot
thick disks with low accretion having polar and disk outflows, and cold
thin disks with high accretion and weak polar and disk outflows.  

Figure~2a shows the flow structure interior to the Bondi radius for
the hot disk with $\dot{M}_{\rm B}/\dot{M}_{\rm Edd}=10^{-4}$.
Red contours are logarithmic contours of density, blue vectors
represent mass flux in the $R-z$ plane, and the solid black line
indicates the zero density surface in the adiabatic settling solution.
Gas flows inwards over a wide range of intermediate polar angles, and
there are equatorial and polar outflows.  The mass ejected in the
polar outflow, $\sim 0.02\dot{M}_{\rm B}$, is similar in magnitude
to the adiabatic case, and is again caused by inflowing matter
turning around exterior to the centrifugal torus.  The detailed properties of this
circulation may depend on the length and accuracy of our numerical
integrations.  The equatorial outflow is driven 
primarily by azimuthal stresses associated with the shear viscosity, and it has magnitude $\sim
0.25\dot{M}_{\rm B}$ at the Bondi radius.  The mass inflow rate
through the Bondi radius increases over that in the inviscid case and
essentially balances this equatorial outflow.  Figure~2b shows red
contours of specific angular momentum in units of
$j_\infty=\sqrt{0.02}R_{\rm B}c_{s,\infty}$, with blue vectors
indicating the flux of angular momentum.  The purple contour shows
where $j=j_\infty$; equatorward the gas has higher specific angular
momentum, and poleward the gas has lower specific angular momentum.
The equatorial outflow carries off the angular momentum of the gas
that was accreted \citep{kolykhalov79,inogamov10} and some of the
angular momentum lost by the polar outflow.

Figure~3a shows logarithmic density contours and velocity vectors
interior to $0.1R_{\rm B}$.  The black line indicates the zero density
surface in the stationary solution.  Inflowing matter dumps angular
momentum into the equatorial regions, and an outflow is driven
beginning just beyond the centrifugal radius, located at $R_{\rm
c}=0.02R_{\rm B}$.  The inflowing gas is able to penetrate below the
centrifugal radius only due to the angular momentum transport,
i.e. the assumed weak MRI induced viscosity.  In Figure~3b, which
shows logarithmic density contours and velocity vectors interior to
half the centrifugal radius, we see that the gas accretes over a range
of latitudes from a time-dependent, vertically extended, sub-Keplerian
disk.  The accretion rate is very low, however, with magnitude $\sim 4
\times 10^{-3}\dot{M}_{\rm B}$ of order $\alpha$ times the mass inflow
rate.  There is some accretion due to the angular momentum transport,
but the bulk of the gas is too energetic, with
$Be>c_{s,\infty}^2/(\gamma-1)$, and has too much angular momentum for
much of it to accrete.  This conclusion is in close agreement with that
of \cite{stone99}.

\begin{figure*}[htp]
\centering
\includegraphics[width=.87\textwidth]{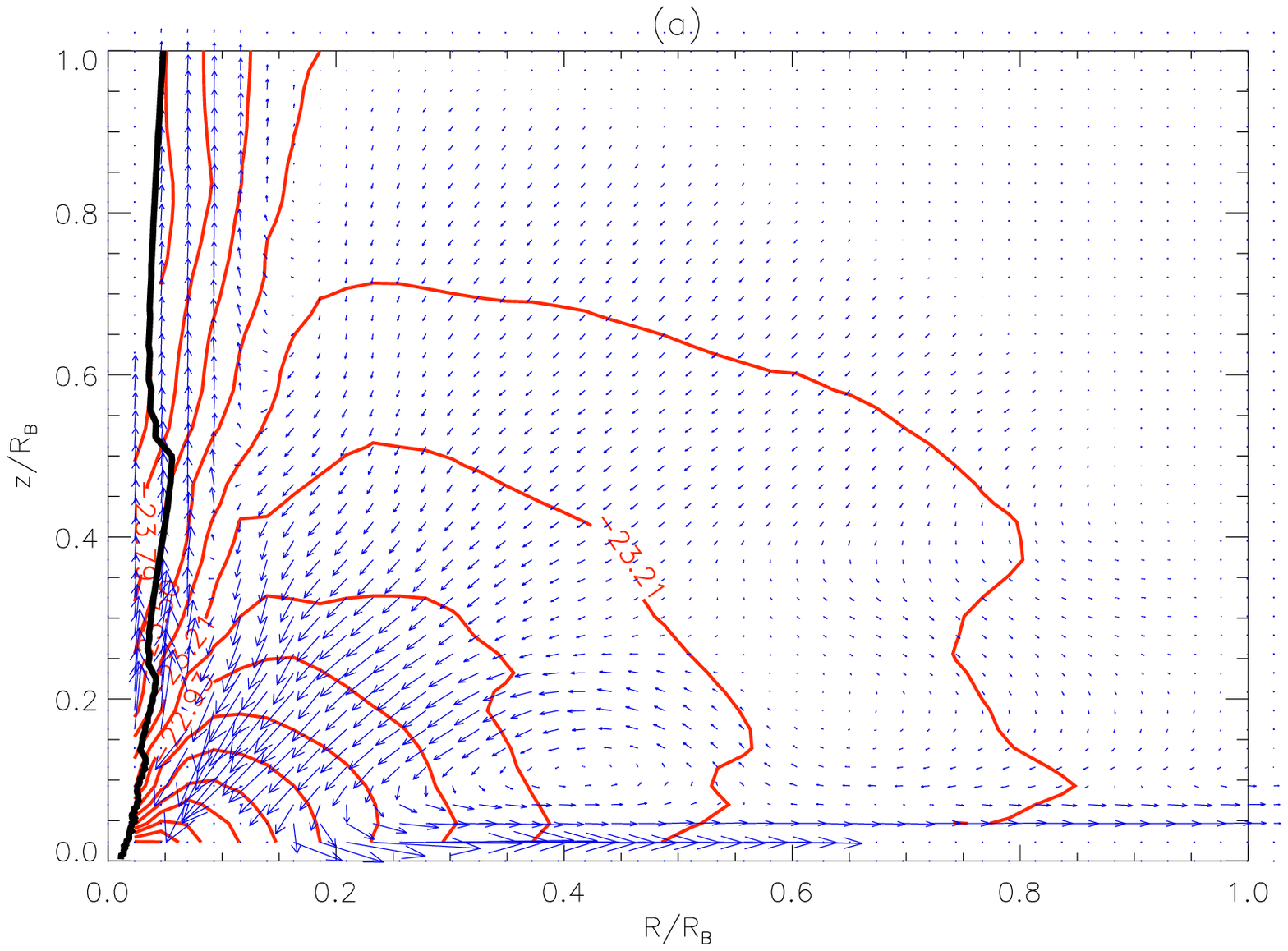}\\
\includegraphics[width=.87\textwidth]{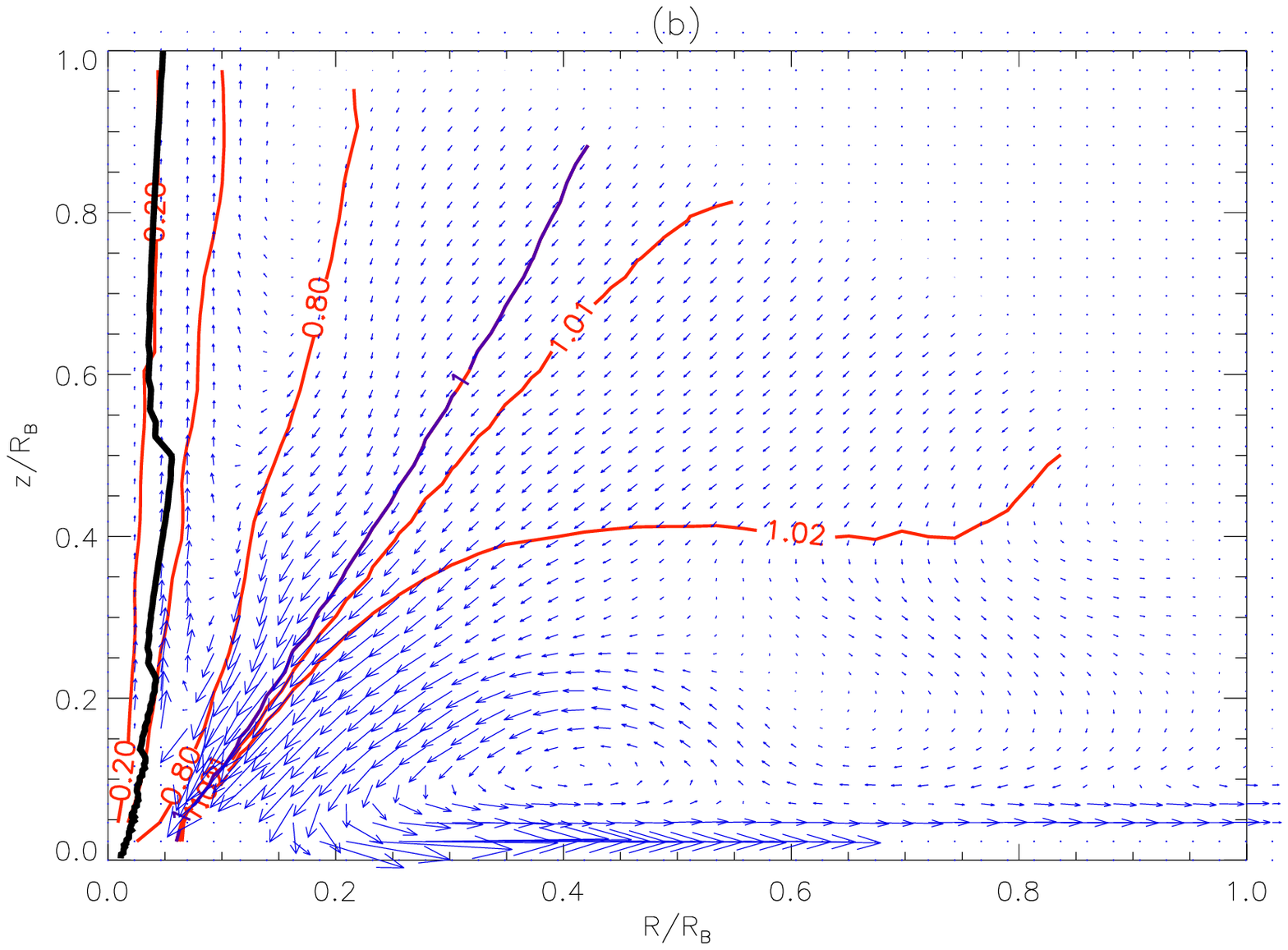}
\caption{(a) The structure of the hot disk with angular momentum
transport designed to mimic the MRI, shown interior to $R_{\rm B}$
in the low accretion limit where $\dot{M}_{\rm B}/\dot{M}_{\rm
Edd}=10^{-4}$.  Cooling is negligible.  Red contours are logarithmic
contours of density, blue arrows indicate mass flux in the $R-z$
plane, and the black line demarcates the zero density surface in the
adiabatic settling solution.  Matter flows in through the Bondi radius
and back out in disk and conical polar outflows.  (b) Contours of specific
angular momentum in units of $j_\infty=\sqrt{0.02}R_{\rm
B}c_{s,\infty}$ and blue vectors of angular momentum flux.  The
black line demarcates the zero density surface in the stationary
solution, and the purple contour marks where the specific angular
momentum $j=j_\infty$.  Angular momentum of inflowing material is
transferred to the equatorial outflow.}\label{fig:viscoushot}
\end{figure*}

\begin{figure*}[htp]
\centering
\includegraphics[width=.87\textwidth]{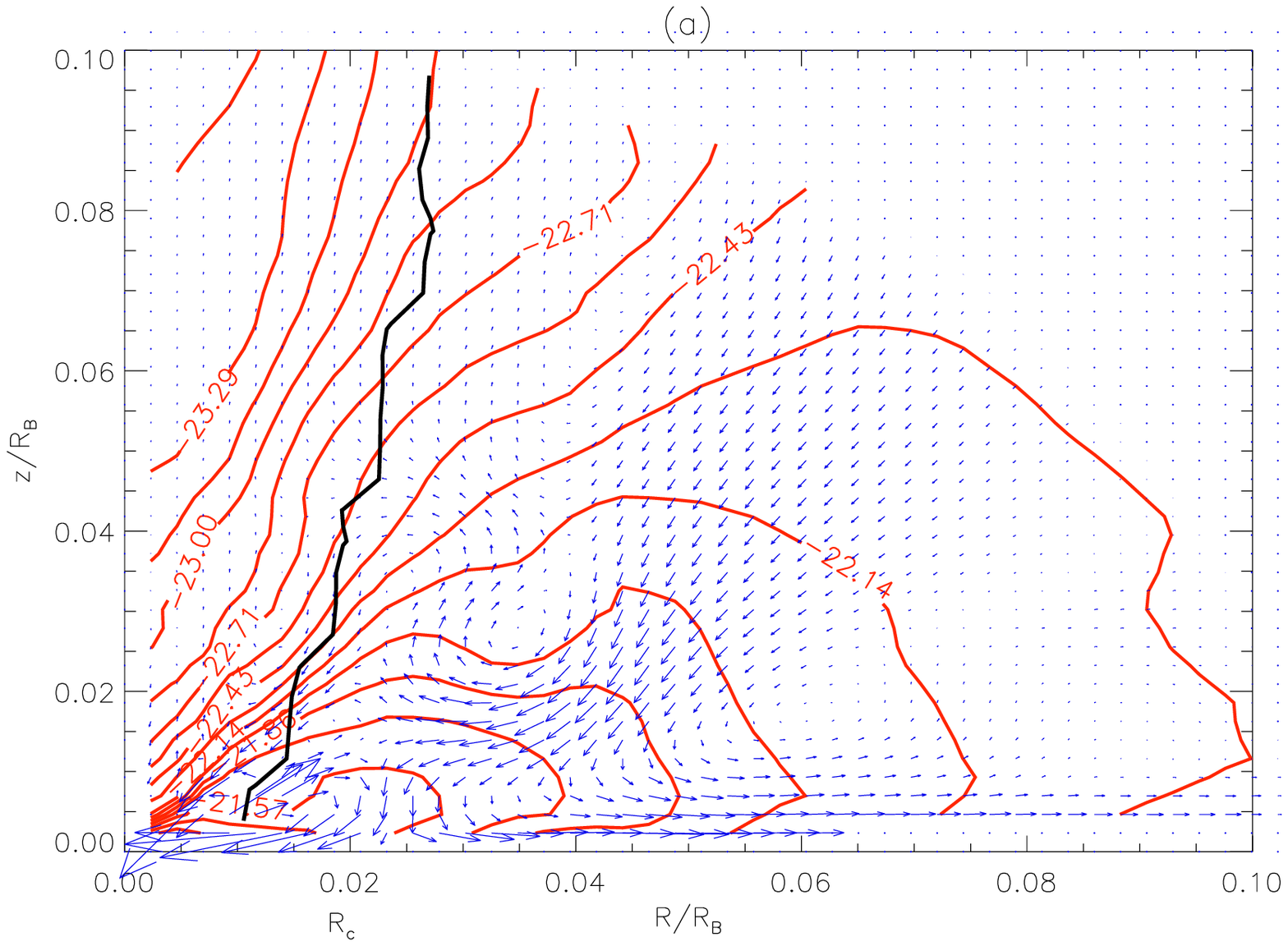}
\includegraphics[width=.87\textwidth]{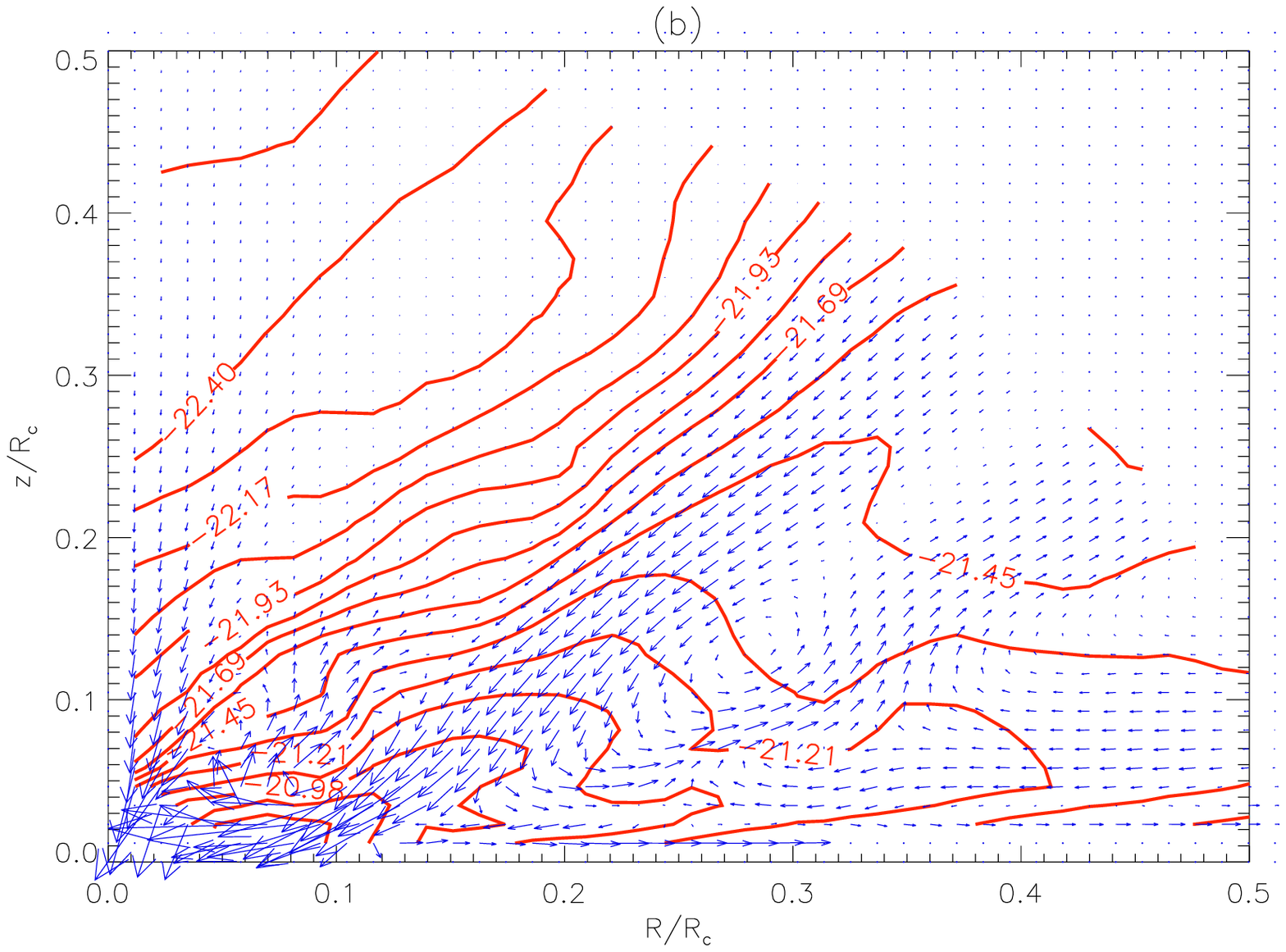}
\caption{Closeup of the structure of the hot disk with angular
momentum transport at Eddington ratio $\dot{M}_{\rm
B}/\dot{M}_{\rm Edd}=10^{-4}$.  This is the same model as
Figure~2.  Red contours are logarithmic contours of density, and blue
arrows represent mass flux in the $R-z$ plane.  (a) Flow structure
interior to $5R_{\rm c}=0.1R_{\rm B}$.  The black line shows
the zero density surface in the adiabatic settling solution.  Angular
momentum transport allows a small amount of matter to inflow below the
centrifugal radius, and viscous stresses drive a strong equatorial
outflow.  (b) Flow structure interior to half the centrifugal radius or
$0.01R_{\rm B}$.  Accretion proceeds at very low rates over a
range of polar angles.}\label{fig:viscoushotsmall}
\end{figure*}

In the cold disk solutions, the non-adiabatic gas radiates energy and
most of the gas cannot travel back out to infinity on energetic
grounds since $Be<c_{s,\infty}^2/(\gamma-1)$.  In the absence of angular momentum transport the gas
continuously accumulates near the centrifugal barrier.  Gas pressure
is insufficient to drive matter inward onto the central black hole and
there is no steady solution.  The introduction of angular momentum
transport allows for significant accretion from the centrifugal torus,
however.  Figure~4a shows logarithmic density contours and mass flux
vectors for the cold disk solution at Eddington ratio $\dot{M}_{\rm
B}/\dot{M}_{\rm Edd}=0.1$, interior to the Bondi radius.  The
solid black line shows the zero density surface in the adiabatic
settling solution given in equation (\ref{eq:dens}).  Matter
accumulates in an extended torus whose pressure slows mass inflow,
similar in spirit to the adiabatic settling solution.  Inflow proceeds
at low polar angles $\theta\lesssim \pi/4$, however, with total inflow
rate $\sim 0.45\dot{M}_{\rm B}$ at the Bondi radius.  Equatorial
gas exterior to the centrifugal barrier is still driven outwards by
the viscous engine, but the outward propagation of the equatorial
outflow is slowed by the strong radiative cooling.  In the absence of
external pressure, the equatorial outflow would continue to propagate
outwards indefinitely \citep{kolykhalov80}.  Our constant pressure
outer boundary condition slows the outflow, however, and we find that
the equatorial outflow is unable to propagate to the Bondi radius on
time scales of order $\sim 100\,t_{\rm B}$.  This solution
is not technically a steady-state solution, as angular momentum is
continuously dumped into the radially extended torus, which acts as a
large sink of angular momentum.  If we integrated for much longer times,
however, the equatorial outflow must continue to propagate outwards to conserve momentum in a steady-state.
Figure~4b shows logarithmic density contours and mass fluxes interior
to $0.1R_{\rm B}$.  The flow forms a geometrically thin disk, and
angular momentum transport allows matter to continue to flow inwards
from this disk.  We illustrate the cold thin disk near the centrifugal
barrier in greater detail in Figure 5.  Red contours are again
logarithmic density contours and blue vectors represent mass flux.
Accretion proceeds through the inner boundary from the cold, thin, sub-Keplerian disk,
at a rate nearly equal to the mass inflow rate through the Bondi
radius.  This case resembles the standard thin disk \citep{shakura73}.

\begin{figure*}[htp]
\centering
\includegraphics[width=.87\textwidth]{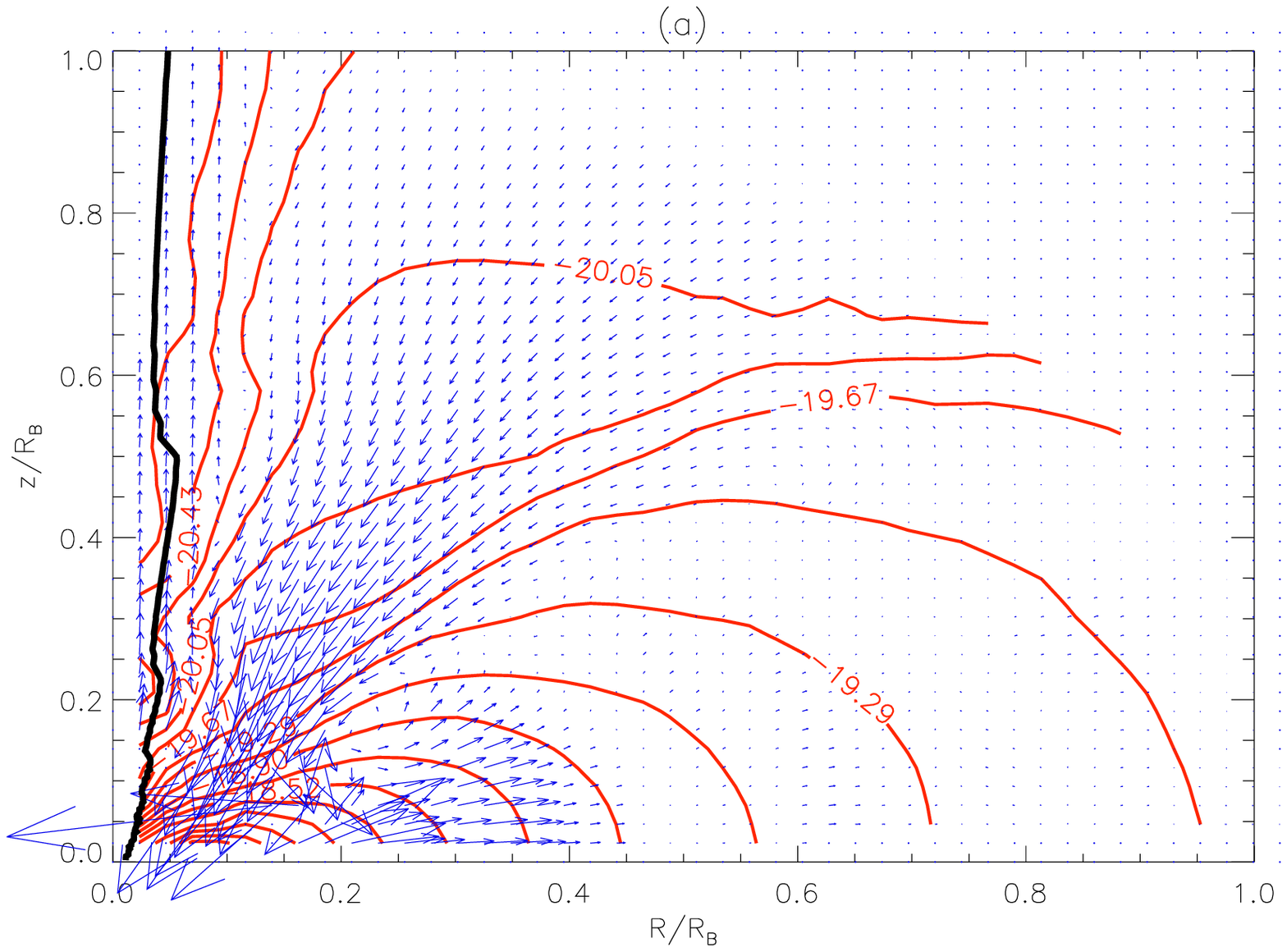}\\
\includegraphics[width=.87\textwidth]{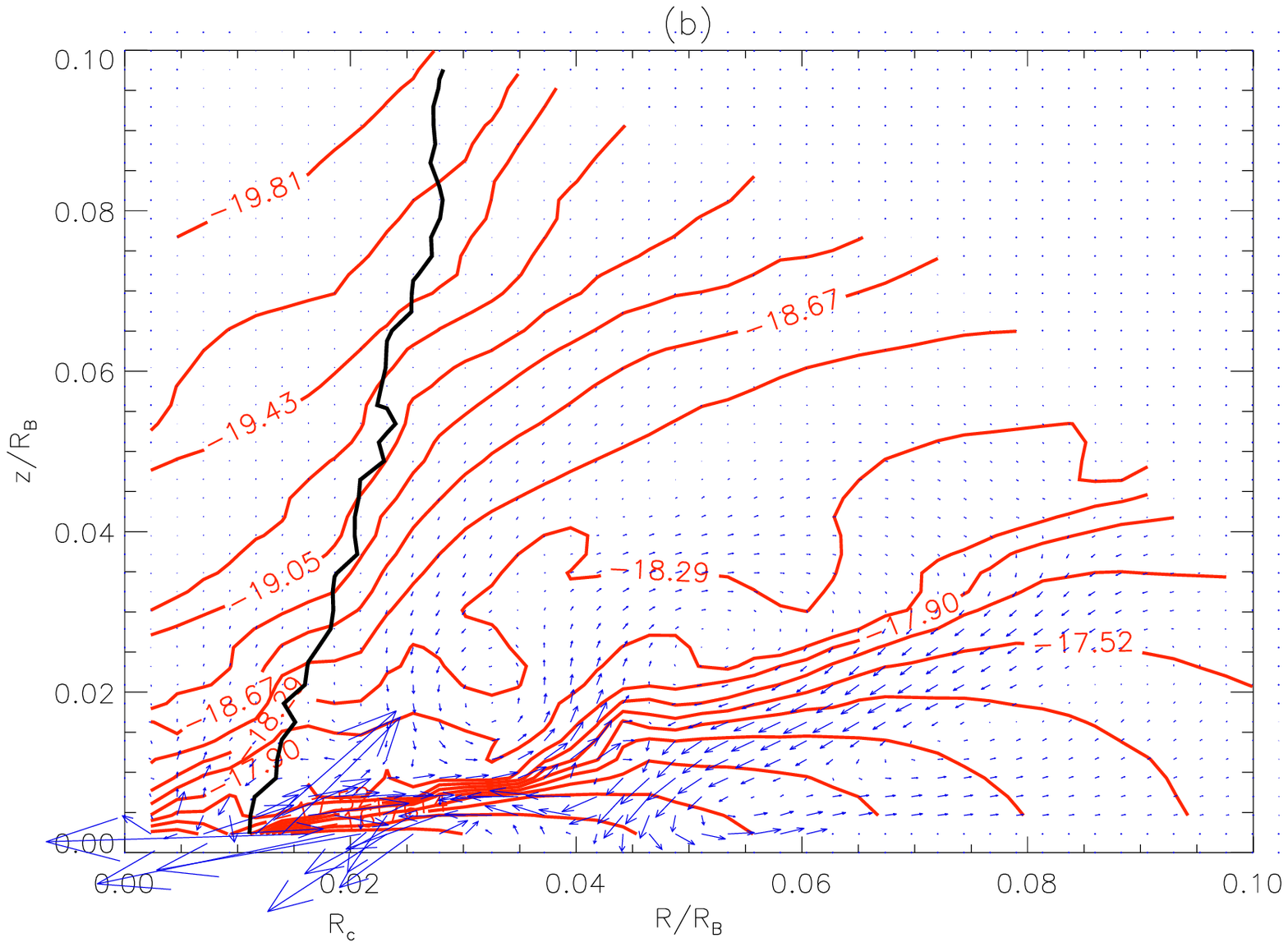}\\
\caption{(a) The structure of the cold disk with angular momentum
transport interior to $R_{\rm B}$, at Eddington ratio
$\dot{M}_{\rm B}/\dot{M}_{\rm Edd}=0.1$.  Red contours are
logarithmic contours of density, and blue arrows indicate mass flux in
the $R-z$ plane.  Inflow is impeded by gas pressure at large polar
angles but proceeds through the Bondi radius at polar angles $\theta
\lesssim \pi/4$.  Angular momentum is dumped into the extended torus.
(b) Flow structure interior to $5R_{\rm c}=0.1R_{\rm B}$.
Matter continues to flow inwards from a geometrically thin disk.  This
is the case which most closely approximates the standard
Shakura-Sunyaev disk.}\label{fig:viscouscold}
\end{figure*}

\begin{figure*}[htp]
\centering
\includegraphics[width=.87\textwidth]{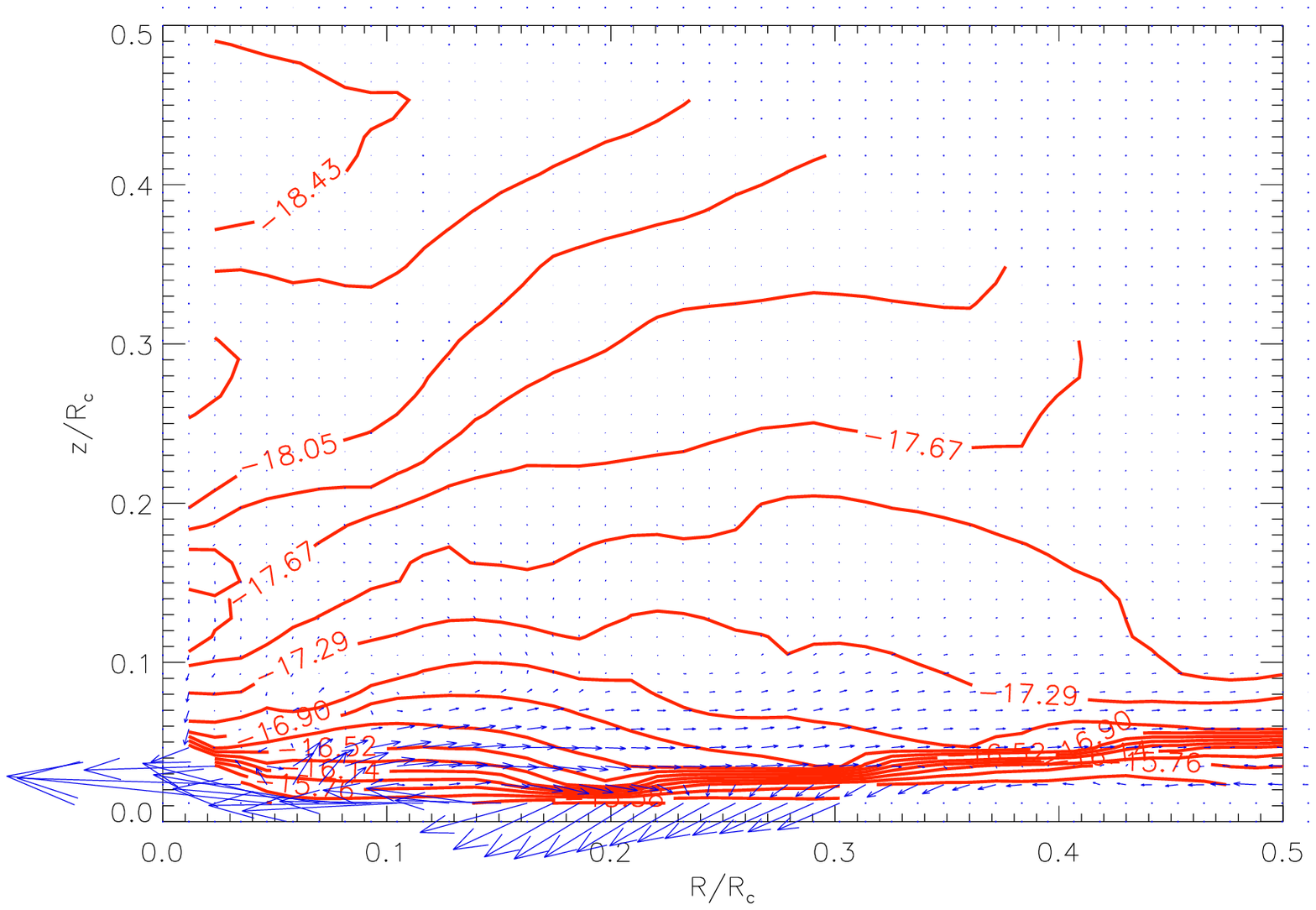}
\caption{A closeup view of the high inflow case with $\dot{M}_{\rm
B}/\dot{M}_{\rm Edd}=0.1$.  We show logarithmic density contours
and mass flux vectors for the cold disk with angular momentum
transport, interior to half the centrifugal radius.  Matter accretes
from the thin disk as in the standard \cite{shakura73} picture with
angular momentum transport due to the MRI.}\label{fig:viscouscoldtiny}
\end{figure*}

\subsection{Radiative Cooling}
\label{sec:radcool}
\begin{figure*}[thbp]
\centering
\includegraphics[width=1.\textwidth]{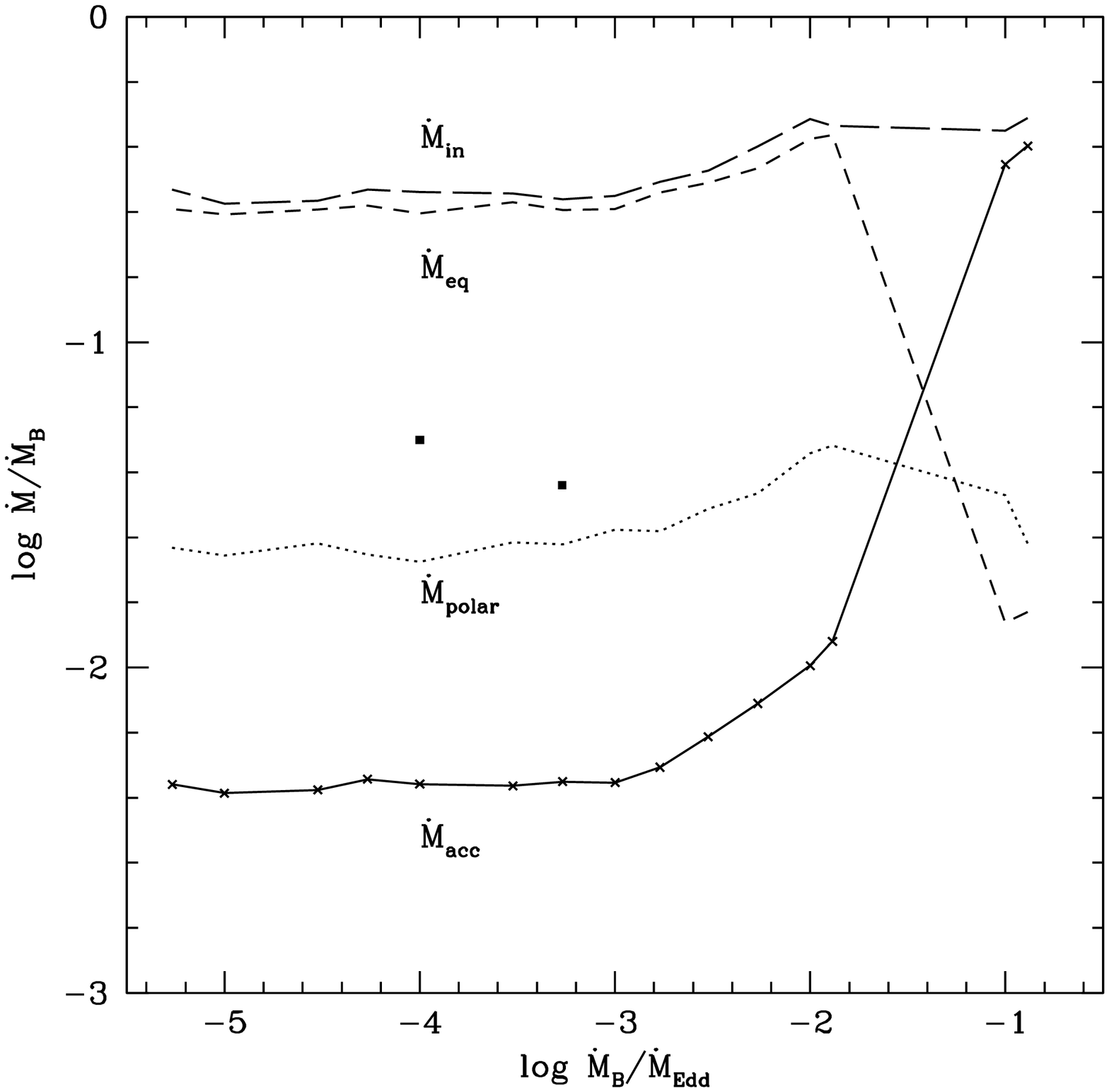}
\caption{Mass inflow rate from large radii, equatorial outflow rate,
polar outflow rate (boxes are averages that include episodes of convective overturning), and
accretion rate as a function of the Eddington parameter $\dot{M}_{\rm
B}/\dot{M}_{\rm Edd}$.  At high $\dot{M}_{\rm B}/\dot{M}_{\rm Edd}$
mass inflows at $\sim1/2$ the Bondi rate and nearly all of it is
accreted.  At low $\dot{M}_{\rm B}/\dot{M}_{\rm Edd}$, mass again
inflows at $\sim1/2$ Bondi, but nearly all of this matter is deflected
outwards in disk outflows.  There is an abrupt transition between
these distinct solutions in the range $\dot{M}_{\rm B}/\dot{M}_{\rm
Edd}\sim 0.01 - 0.1$.}\label{fig:accretion_wind}
\end{figure*}

Figure~\ref{fig:accretion_wind} shows the quasi-steady mass accretion
rate through the inner boundary, the equatorial and polar wind mass
outflow rates through the Bondi radius, and the mass inflow rate
through the Bondi radius, all as functions of the parameter
$\dot{M}_{\rm B}/\dot{M}_{\rm Edd}$.  Crosses on the accretion rate
curve indicate the Eddington ratios for which our simulations have
converged with respect to temporal duration.  For $\dot{M}_{\rm
B}/\dot{M}_{\rm Edd}\lesssim 0.01$, the Bremsstrahlung cooling is weak
and plays little dynamic role in the flow.  The flow converges to a
low accretion rate solution for $\dot{M}_{\rm B}/\dot{M}_{\rm
Edd}\lesssim 10^{-3}$.  The non-zero viscosity allows only a small
fraction, $\sim 0.4$\%, of the inflowing matter at the Bondi radius to
accrete onto the central black hole.  The mass inflow rate at the
Bondi radius $\dot{M}_{\rm inflow}$ is itself only $\sim 0.29
\dot{M}_{\rm B}$ because the centrifugal torus and outflows
significantly impede the inflow (cf Figures 2 and 3).  The gas density
on the equator is much larger than in the polar region, and the
equatorial outflow dominates the outward mass and momentum flux at the
Bondi radius.  The equatorial outflow nearly matches the mass inflow
rate at the Bondi radius, whereas the polar outflow is much less
significant and an order of magnitude weaker than either.  It is
important to note, however, that viscous heating near the centrifugal
radius leads to a buildup of thermal energy and can lead to
intermittent episodes of convective overturning.  The overturning
episodes can last for $\sim10 t_B$, with outward mass fluxes over a
broad range of polar angles and a significant fraction of the
characteristic mass inflow rates.  The filled boxes in
Figure~\ref{fig:accretion_wind} represent estimates of the
time-averaged polar mass outflows including these overturning
episodes, which by definition are not included in our quasi-steady
average mass fluxes (see Appendix for more detail).

The strength of the radiative cooling increases with the
parameter $\dot{M}_{\rm B}/\dot{M}_{\rm Edd}$, and in the range
$\dot{M}_{\rm B}/\dot{M}_{\rm Edd}\sim 0.01 - 0.1 $ there is an abrupt
transition in the properties of the flow to the strong cooling regime.
For $\dot{M}_{\rm B}/\dot{M}_{\rm Edd}\sim 0.1$ the inflowing gas is
impeded in the equatorial regions by gas pressure in an extended
centrifugal torus (cf Figure 4), and matter flows in through the Bondi
radius at roughly half the Bondi rate.  Nearly the entirety of this
inflowing gas is accreted onto the central black hole and the
solutions approach the standard \cite{shakura73} picture.  Most of the
inflowing gas does not have enough energy to travel back out to
infinity, and only a very small portion bounces off the centrifugal
barrier and is ejected in a polar wind.  The persistence of the polar
wind is uncertain, and much higher resolution studies with additional
physics such as magnetic fields and radiation are necessary to study
the detailed properties of polar outflows.  We emphasize again that
the high Eddington ratio solutions are only quasi-steady (see Section
\ref{sec:hotcold}), and the equatorial outflows would be stronger if
we integrated for times $\gtrsim 1000 R_{\rm B}/c_{s,\infty}$.

\begin{figure*}[thbp]
\centering
\includegraphics[width=.5\textwidth]{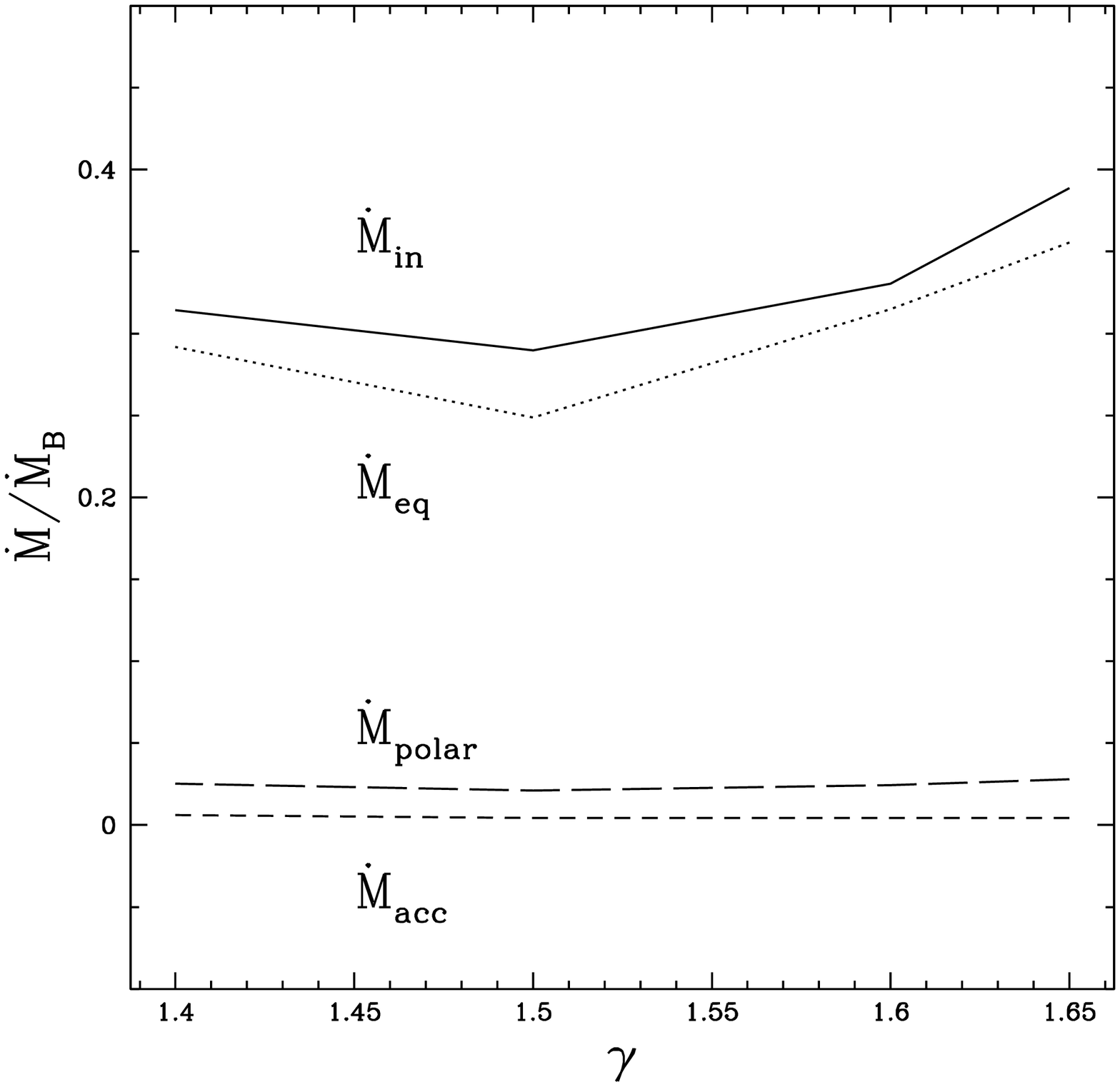}
\caption{Mass inflow rate from large radii, equatorial and polar
outflow rates at $R_{\rm B}$, and accretion rate as a function of
adiabatic index $\gamma$.  All solutions are computed for fixed
Eddington ratio $\dot{M}_{\rm B}/\dot{M}_{\rm Edd}=10^{-4}$, the
hot disk solutions.  Mass fluxes have relatively weak dependence on
adiabatic index in the range illustrated, and we set $\gamma=1.5$
throughout the rest of this work.}\label{fig:gamma}
\end{figure*}

We have further verified that the low accretion rate solutions depend
weakly on our choice of adiabatic index.  Figure~\ref{fig:gamma} shows
the mass inflow rate from large radii, equatorial and polar outflow rates at $R_{\rm B}$, and accretion rate as a function of
$\gamma$.  Mass fluxes depend weakly on $\gamma$ in the range
$1.4-1.65$, and our ansatz $\gamma=1.5$ should not significantly
affect our results.

The high accretion rate flows from Figure~\ref{fig:accretion_wind}
with $\dot{M}_{\rm B}/\dot{M}_{\rm Edd}>0.1$ are cold and form a
geometrically thin disk, and the low accretion rate flows with
$\dot{M}_{\rm B}/\dot{M}_{\rm Edd}<0.01$ are hot and form
vertically extended disks.  Figure~8a shows the emission weighted
temperature,
\begin{equation}
<T>\equiv \frac{\int \dot{e}_{\rm Brem} T dV}{\int\dot{e}_{\rm Brem} dV},
\end{equation}
as a function of $\dot{M}_{\rm B}/\dot{M}_{\rm Edd}$.  We show
three characteristic temperature curves, from top to bottom averaged
over the region interior to the centrifugal barrier, the total region
within $0.1R_{\rm B}$, and in the centrifugal torus.  The
centrifugal torus is extremely dense and cools to a lower temperature
(the floor that we set) than other parts of the flow at high Eddington
ratios.  The gas interior to the centrifugal barrier is hottest
because the gas has penetrated deeper into the gravitational potential
well.  For low $\dot{M}_{\rm B}/\dot{M}_{\rm Edd}<0.01$ the gas in
the inner parts of the flow is hot and a sizable fraction of the
centrifugal temperature
\begin{equation}
T_{\rm c}\equiv \frac{\mu m_p}{k_b \gamma}\frac{2 GM_{\rm BH}}{ R_{\rm c}}=100T_\infty,
\end{equation}
where $\mu=0.5$ is the mean molecular weight.  
Electrons become weakly
relativistic at such temperatures, but we neglect in our computations
the increase of the Bremsstrahlung energy losses for relativistic
electrons.  
We measure the pressure scale height of the disk as the
vertical distance at the centrifugal radius over which the pressure
drops by a factor of $e$.  The pressure scale height of the disk for
the hot solutions with vertically extended disks is $H_p/R_{\rm c}\sim
0.75$ (see Figure~8b), and they have $\alpha\sim0.01$ at the
centrifugal barrier.  The high accretion rate solutions with cold thin
disks have pressure scale at the centrifugal radius $H_p/R_{\rm
c}\sim0.15$, covering $\sim 5$ cells in the highest resolution
simulation (see Appendix).  For a cold thin disk in hydrostatic
equilibrium and with angular momentum $j=\sqrt{0.02}R_{\rm
B}c_{s,\infty}$ we expect $H_p/R_{\rm c}=c_s/v_\phi=0.07$.  Since we
fix the kinematic viscosity $\nu$ in our code, our cold disks have
$\alpha\sim0.4$ rather than $\alpha\sim0.01$ at the centrifugal
barrier.  The code has difficulty running the cold solutions at lower
viscosity, as there is significantly greater mass deposited near the
centrifugal radius before accretion can proceed.  The gas has cooled
strongly and energetically cannot reach infinity, however, and we
expect it to accrete independently of the strength of the viscosity,
as long as it is non-zero.  In any case, radiative transfer becomes
important at Eddington ratios of $\sim {\rm few} \times 0.01$, physics
that we do not include here.  The precise location of the transition
between thin and thick disks, as well as the density, luminosity, and
optical depth of the thin disks, will depend on these choices.

The conical polar outflows, driven by inflowing matter bouncing off
the centrifugal barrier, are in general weak, but they exist over a
wide range of Eddington ratios.  Figure~8c gives the half-opening
angle of the polar outflow at the Bondi radius as a function of
$\dot{M}_{\rm B}/\dot{M}_{\rm Edd}$.  The half-opening angle is
typically $\theta_{\rm half-open}\sim 5^{\circ}$.  We show also the
half-angles at which half of the mass and half of the momentum in the
polar outflows are ejected, again computed at the Bondi radius.  The
zero density surface in the adiabatic settling solution lies at an
angle of $\sim 3^{\circ}$ at the Bondi radius.  Figure~8d shows the
mass-flux weighted average wind velocity in the polar outflow,
measured at the Bondi radius, as a function of $\dot{M}_{\rm
B}/\dot{M}_{\rm Edd}$.  The wind velocity can be a significant
fraction of the centrifugal velocity
\begin{equation}
v_{\rm c}\equiv (2GM_{\rm BH}/R_{\rm c})^{1/2},
\end{equation}
with typical wind velocity $<v_{\rm wind}>\sim 0.15\, v_{\rm c}$.  This value is similar to that found in \citep{yuan12b}.  The outflow opening angle and velocity
depend on the radius at which we measure these quantities.  The
outflow propagates into a constant density and pressure medium in our
setup, and at sufficiently large distances the outflow will transfer
its momentum and energy to the external medium.  Beyond this region
the outflow would become subsonic and probably circulate back to join
the inflowing gas at intermediate latitudes.  We quote values at the
Bondi radius to give a quantitative estimate for the outflow
properties at a definite radius.

We now discuss in greater detail the radiative properties of our
accretion flows.  If a spherically symmetric distribution of gas is in
free-fall with velocity
\begin{equation}
v_{\rm ff}=(2GM_{\rm BH}/r)^{1/2}
\end{equation}
and number density
\begin{equation}
n(r)=\frac{\dot{M}_{\rm B}}{4\pi c R_s^2 \eta}(r/R_s)^{-3/2},
\end{equation}
where $\eta$ is the mean mass per electron, then the optical depth
\begin{equation}
\tau\equiv \sigma_T \int n(r) dr \propto \dot{M}_{\rm B}/\dot{M}_{\rm Edd}.
\end{equation}
We compute the optical depth along three different sightlines with
fixed polar angle $\theta=0^{\circ},45^{\circ},90^{\circ}$, starting
from the inner boundary at $R_{\rm in}=91R_s$.  We find that $\tau$
scales linearly with $\dot{M}_{\rm B}/\dot{M}_{\rm Edd}$ (see
Figure~8e).  The optical depths approach $10^{-2}$ just below $\dot{M}_{\rm
B}/\dot{M}_{\rm Edd}=10^{-2}$, where the flow switches between
the hot, vertically extended disk and the cold, thin disk, and there is a jump by a
factor of $\sim 100$ in $\tau_{\rm 90}$ at this transition.  Further,
making use of the temperature profile of an adiabatic gas in
free-fall,
\begin{equation}
T\simeq T(R_s)(r/R_s)^{-3(\gamma-1)/2},
\end{equation}
the total integrated Bremsstrahlung luminosity
\begin{align}
L_{\rm Brem}/L_{\rm Edd} &\propto \frac{1}{L_{\rm Edd}}\int n(r)^2
T^{1/2} r^2 dr \\ &\propto (\dot{M}_{\rm B}/\dot{M}_{\rm Edd})^2.
\end{align}
Figure~8f shows the Bremsstrahlung luminosity in our simulations
integrated over the regions between $R_{\rm in}=91R_s$ and $0.1R_{\rm
B}$, and between $R_{\rm in}=91R_s$ and the centrifugal torus.  We
obtain the scaling $L_{\rm Brem}/L_{\rm Edd} \propto (\dot{M}_{\rm
B}/\dot{M}_{\rm Edd})^2$ for both integrated luminosities, and $L_{\rm
Brem}\to 10^{-7}L_{\rm Edd}$ at $\dot{M}_{\rm B}/\dot{M}_{\rm
Edd}=10^{-2}$.  As we move the inner boundary at $R_{\rm in}$ closer
to the black hole, we expect the luminosity, temperature, and optical
depth interior to the centrifugal radius to increase.  Other work
including general relativistic and magnetohydrodynamic physics and
extending down to the horizon for non-spinning black holes suggests that essentially all matter flowing inwards at our inner boundary $R_{\rm in}=91 R_s$ will ultimately accrete,
however \citep{narayan12}.

\begin{figure*}[htp]
\centering
\includegraphics[width=1.\textwidth]{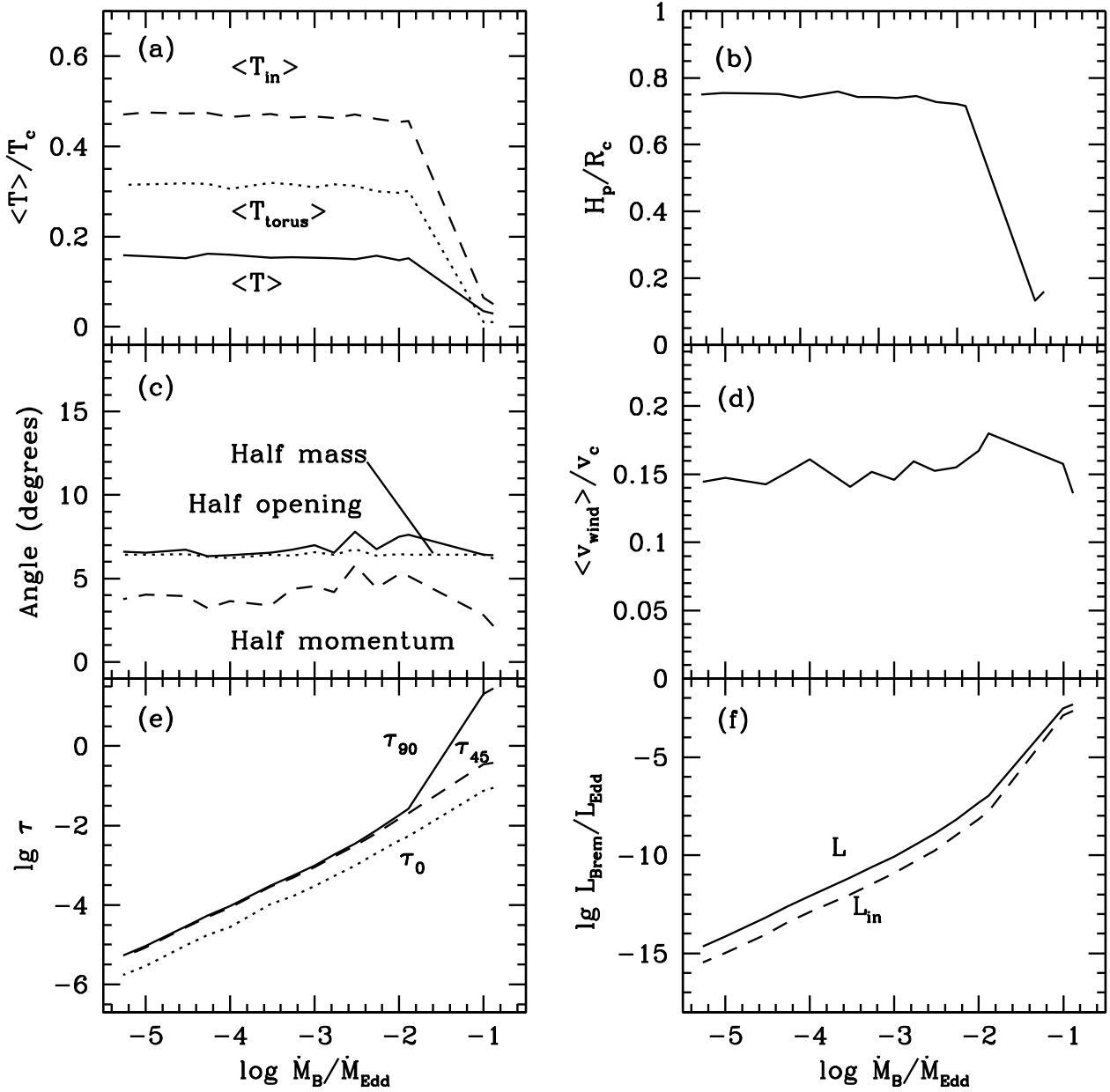}\\
\caption{6 characteristic quantities in our solutions, all as a
function of $\dot{M}_{\rm B}/\dot{M}_{\rm Edd}$.  (a) Emission
weighted temperature, averaged over the region interior to the
centrifugal torus, the region interior to $0.1 R_{\rm B}$, and the
centrifugal torus.  The low accretion rate solutions are hot, with
temperature a sizable fraction of the centrifugal temperature,
whereas the high accretion rate solutions are cold.  (b) Pressure
scale height.  The hot disks are vertically extended, with $H_p/R_{\rm
c}\sim 0.75$, whereas the cold disks are geometrically thin, with
$H_p/R_{\rm c}\ll 1$.  (c) Polar wind half-opening angle, measured
at $R_{\rm B}$, typically of order $5^{\circ}$.  We show also the
half-angles at which half the mass and momentum in the wind are
ejected.  (d) Mass-flux weighted polar wind velocity, measured at
$R_{\rm B}$, typically a sizable fraction of the centrifugal
velocity.  (e) Optical depth along three sightlines with fixed
$\theta=0^{\circ},45^{\circ},90^{\circ}$.  $\tau$ scales linearly with
$\dot{M}_{\rm B}/\dot{M}_{\rm Edd}$, but there is a jump in
$\tau_{90}$ when the disk switches to a dense, cold thin disk.  The
optical depths approach $10^{-2}$ at $\dot{M}_{\rm B}/\dot{M}_{\rm
Edd}=10^{-2}$.  (f) Total Bremsstrahlung luminosity integrated over
the region between the inner boundary and $0.1R_{\rm B}$, and
between the inner boundary and the centrifugal torus.  Luminosity
scales as the second power of $\dot{M}_{\rm B}/\dot{M}_{\rm Edd}$
and approaches $10^{-7}L_{\rm Edd}$ at $\dot{M}_{\rm
B}/\dot{M}_{\rm Edd}=10^{-2}$.}\label{fig:mdot}
\end{figure*}

\section{Discussion}\label{sec:discussion}
Observations suggest that the Active Galactic Nuclei (AGN) in nature
form a bimodal distribution, with broad-line AGN having accretion
rates $\sim 2$ orders of magnitude higher than narrow-line and
lineless AGN \citep{trump09,trump11,kollmeier06,russell12}.  The transition
between these classes occurs at an Eddington ratio between $0.01$ and
$0.1$, in remarkable agreement with our transition between hot and
cold flows.  Further, the luminosities of the cold and hot disk states
suggests a natural feedback loop that can result in switching between
the two solutions.  From Figure~\ref{fig:mdot} we see that the
Bremsstrahlung emission from a cold thin disk can approach
$10^{-2}L_{\rm Edd}$ for $\dot{M}_{\rm B}/\dot{M}_{\rm Edd}\sim0.1$.
We do not include the radiative forcing or thermal heating effects of
this emission on infalling gas in our code
\citep{park01,park07,yuan09}, but the Bremsstrahlung luminosity would
significantly heat the inflowing matter.  The disk could transition to
the hot, vertically extended state with low accretion, effectively
shutting off the Bremsstrahlung emission.  In the absence of
irradiating photons, the hot disk could then cool and begin accreting
at high rates again, thus forming a cycle.  This process may be an
important detail in explaining the high and low accretion states of
quasars \citep{ciotti01}.  The physics of radiative feedback on
supermassive black hole accretion flows has been studied numerically
in some detail
\citep{sazonov05,proga07,nagamine11,ciotti09,ciotti10,kurosawa09,proga08},
and a potentially significant effect here would be to increase the
strength of the polar outflows.

Independently of any radiative feedback loops, our results indicate
that temporal variations in the external boundary conditions at
infinity would lead to strongly time-dependent accretion flows.
Observations are now approaching sufficient accuracy that for nearby
supermassive black holes both the external boundary conditions
surrounding the black hole and the reaction of the flow to these
external boundary conditions can be measured.  Indeed,
\cite{gillessen12} recently discovered a gas cloud of only several
Earth masses orbiting Sgr ${\rm A}^*$ and with radius of closest approach
$\sim 3100 R_s$.  The computations provided here are detailed enough
to make concrete predictions concerning the geometrical, dynamical,
and thermal properties of accretion disks and outflows of nearby
supermassive black holes from currently feasible observations.  We do
not provide detailed predictions here but may further explore these
issues in a separate paper addressing observational consequences.

The lack of magnetic fields is a limitation of our work \citep{pessah08,beckwith09,penna10,tchekhovskoy11,mckinney12}.
\cite{proga03b} considered the non-radiative two-dimensional
magnetohydrodynamic accretion problem, and they find angular momentum
transport and accretion driven by the MRI.  The two-dimensional
structure of the flow determines the accretion properties, with polar
outflows quenching the accretion rate.  Similar qualitative behavior
is present in the Adiabatic Inflow Outflow Solutions
\citep{blandford99}, and we also see circulation and polar outflows in
our hydrodynamical simulations with weak cooling.  In principle the
nonlinear properties of the MRI could modify the detailed structure of
the flow in the weak cooling regime.  As the cooling rate increases,
however, the total energy of the fluid determines its fate and the
detailed kinetic properties are less important.

Another extension of our work relates to the enforcement of reflection
symmetry through the equatorial plane.  The flow does not necessarily
need to preserve this symmetry, and in principle there could be
asymmetric, unstable modes that seed turbulence and/or convection that
modifies the properties of the accretion flow.
We have explored this idea in some detail, and we find that the main
difference from the presented results to be that the strong equatorial
outflow of high angular momentum gas need no longer be centered
exactly on the equator, as in \cite{stone99} and \cite{yuan12a}.
Asymmetries in bipolar outflows would have clear observational
consequences as well, and the behavior of these outflows in different
cooling regimes is an avenue of potential research.  In any case
future studies of the global structure of accretion flows should
consider improvements such as fully 3D simulations, general
relativistic corrections, self-gravity, and more detailed heat and
radiation transport mechanisms.  One heat transport mechanism we have
studied in some detail at low Eddington ratios is conduction due to
thermal electrons \citep{inogamov10}.  At lower Prandtl number
convective overturning of hot gas can be suppressed, and the strength
of the inflow and outflow can vary by a factor of a few.  The
distribution of outflow between polar and equatorial outflow also
depends on the strength of the conduction, but the essential feature
of mass accretion rates nearly two orders of magnitude below the Bondi
rate remains unchanged.

\section{Conclusions}
\label{sec:conclusions}
Our high inflow rate solutions are very similar to the standard and
widely adopted \cite{shakura73} results. But our low inflow results
are to zeroth order the stationary \cite{papaloizou84} solution.  To
next order in the small assumed viscosity they show patterns of circulation,
with outflow almost balancing inflow and the net accretion rate
through a geometrically thick disk falling to only the order of
$\alpha$ (the dimensionless viscous parameter) times the inflow
rate. We label this behavior a RRIOS solution for "Radiating, Rotating
Inflow-Outflow Solution", as it is a significant generalization of the
ADIOS (Adiabatic Inflow-Outflow Solution of \cite{blandford99})
applicable to the common case wherein the angular momentum load of the
inflowing gas strongly restricts the level of net accretion, at low
inflow rates.  In this low inflow state, viscous forces can drive an
equatorial outflow that propagates out beyond the Bondi radius.
Further, entropy released by viscous stresses can accumulate near the
centrifugal barrier and lead to episodes of convective overturning.
Observationally, the solutions that we find appear to be consistent
with the fact that most nearby massive black holes show very low
bolometric luminosity, considerably below that anticipated for
spherically symmetric (non-rotating) Bondi-Hoyle flows, but also show
occasional outbursts indicating a much higher rate of episodic
outflow.

We thank Jim Stone for many discussions and for help with the ZEUS
code.  We thank also, Daniel Proga, Feng Yuan, Jim Pringle, Sasha Tchekhovskoy, Yan-Fei Jiang, Bob Penna, Eve Ostriker, Patrick Hall, \&
Anatoly Spitkovsky for useful comments.  The simulations presented in
this article were performed on the Orbital and Della clusters
supported by the PICSciE-OIT High Performance Computing Center and
Visualization Laboratory.

\appendix

\begin{figure}[htp]
\centering
\includegraphics[width=1.\textwidth]{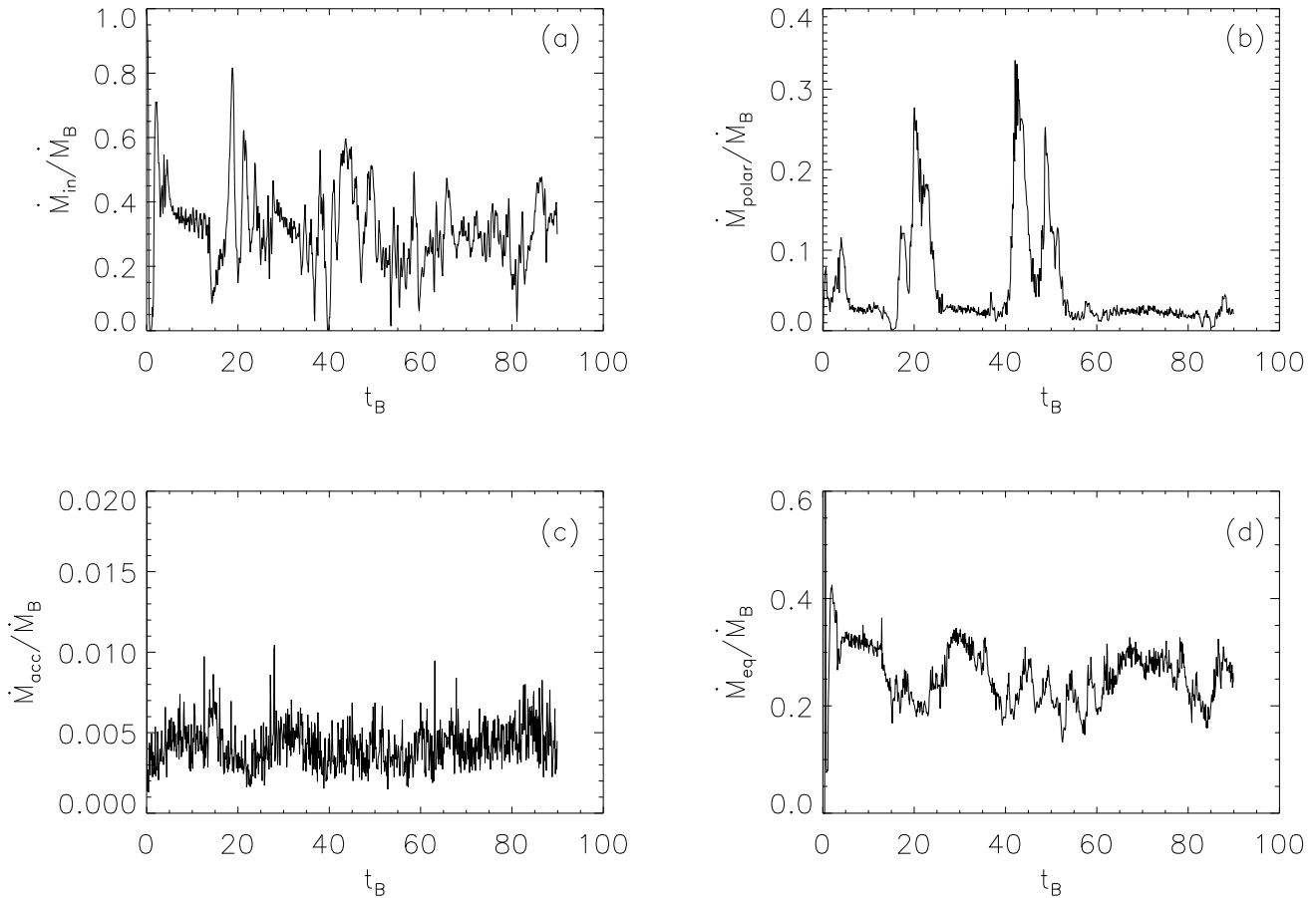}
\caption{Time dependence of mass inflow rate at $R_{\rm B}$,
equatorial and polar outflow rates at $R_{\rm B}$, and the accretion
rate for the solution at $\dot{M}_{\rm B}/\dot{M}_{\rm Edd}=10^{-4}$.  The simulation has converged to a unique steady-state solution after $t\sim 55 t_{\rm B}$.}\label{fig:mdottime}
\end{figure}

In this Appendix we provide further detail on the general properties
of our solutions, including convergence tests.
Figure~\ref{fig:mdottime} illustrates the time dependence of
characteristic quantities in our solutions for Eddington ratio
$\dot{M}_{\rm B}/\dot{M}_{\rm Edd}=10^{-4}$ and computed at our
standard resolutions.  We show the mass inflow rate at $R_{\rm B}$,
equatorial and polar outflow rates at $R_{\rm B}$, and the accretion
rate.  The mass inflow rate, equatorial outflow rate, and accretion
rate are statistically-steady over the illustrated interval.  The
polar outflow at large radii is punctuated by bursts of convective
overturning with high mass outflow rates, but is also quasi-steady
after time $t\sim 55 t_{\rm B}$.  The simulations of, e.g.,
\cite{stone99} and \cite{yuan12a}, found persistent convective eddies
disrupting mass inflow near the centrifugal support radius, but we
find that the thermal energy in the subsonic hot accretion flow
with enforced symmetry through the equator can only support intermittent episodes
of overturning extending to the Bondi radius.  Table \ref{convp0001}
shows the dependence of the four quasi-steady mass fluxes from
Figure~\ref{fig:mdottime} on the spatial resolution of our simulations
for the same Eddington ratio $\dot{M}_{\rm B}/\dot{M}_{\rm
Edd}=10^{-4}$.  The first column gives the factor by which the number
of grid points in each spatial dimension is multiplied.  The accretion
rate in standard resolution simulation is within $5$\% of the value at
double the grid resolution, and the other mass fluxes are within $\sim
30$\%.

\begin{deluxetable}{ccccc}
\tablewidth{0pt}
\tablecaption{Resolution dependence of mass inflow rate at $R_{\rm B}$,
equatorial and polar outflow rates at $R_{\rm B}$, and the accretion
rate for $\dot{M}_{\rm B}/\dot{M}_{\rm Edd}=10^{-4}$.}
\tablehead{\colhead{Resolution} & \colhead{$\dot{M}_{\rm in}/\dot{M}_{\rm B}$} & \colhead{$\dot{M}_{\rm polar}/\dot{M}_{\rm B}$} & \colhead{$\dot{M}_{\rm eq}/\dot{M}_{\rm B}$} & \colhead{$\dot{M}_{\rm acc}/\dot{M}_{\rm B}$}}
\startdata
.81 & 0.32 & 0.034  & 0.29  & 0.0041 \\ \hline
1. & 0.27 & 0.021 & 0.24  &0.0042 \\ \hline
1.25 & 0.27 & 0.020  & 0.25  & 0.0044  \\ \hline
1.5 & 0.24 & 0.020  & 0.22  & 0.0045  \\ \hline
2 & 0.21 & 0.015  & 0.18 & 0.0040

\enddata
\label{convp0001}
\end{deluxetable}

We perform the same exercise at Eddington ratio $\dot{M}_{\rm
B}/\dot{M}_{\rm Edd}=10^{-1}$ and show the quasi-steady average mass
fluxes in Table \ref{convp1}.  The mass inflow rate at $R_{\rm B}$,
polar outflow rate at $R_{\rm B}$, and the accretion rate for the
standard resolution simulation are within $\sim 25$\% of the values in
the simulation with double the grid resolution.  Note that mass fluxes
into and out of our grid do not have to exactly balance during a
quasi-steady phase, as there can be a net buildup of gas near the
centrifugal radius.  This excess gas will eventually be ejected
outwards during a convective overturning episode.

\begin{deluxetable}{ccccc}
\tablewidth{0pt}
\tablecaption{Resolution dependence of mass inflow rate at $R_{\rm B}$,
equatorial and polar outflow rates at $R_{\rm B}$, and the accretion
rate for $\dot{M}_{\rm B}/\dot{M}_{\rm Edd}=10^{-1}$.}
\tablehead{\colhead{Resolution} & \colhead{$\dot{M}_{\rm in}/\dot{M}_{\rm B}$} & \colhead{$\dot{M}_{\rm polar}/\dot{M}_{\rm B}$} & \colhead{$\dot{M}_{\rm eq}/\dot{M}_{\rm B}$} & \colhead{$\dot{M}_{\rm acc}/\dot{M}_{\rm B}$}}
\startdata
.81 & 0.51 & 0.061  & 0.004  & 0.59 \\ \hline
1. & 0.45 & 0.034 & 0.014  &0.35 \\ \hline
1.25 & 0.49 & 0.033  & 0.006  & 0.40  \\ \hline
1.5 & 0.55 & 0.040  & 0.001  & 0.48  \\ \hline
2 & 0.59 & 0.044  & 0.001 & 0.47

\enddata
\label{convp1}
\end{deluxetable}

We also explored the dependence of our solutions on our choice of no
torque inner boundary condition located at $R_{\rm in}=91R_s$.  We
have determined that mass fluxes at large radii change by a factor of
$1.1$, mass fluxes in the polar region at large radii increase by a
factor of $1.2$, and accretion rates decrease by a factor of $1.45$,
as we move the inner boundary to $R_{\rm in}=91R_s/2$
for solutions with $\dot{M}_{\rm B}/\dot{M}_{\rm Edd}=10^{-4}$.
Moving the inner boundary inwards by a factor of $4$ to $R_{\rm in}=91R_s/4$ for the same
Eddington ratio reduces the accretion rate by a factor of $2$.
There is certainly a dependence of accretion flow eigenvalues on the
location of the inner boundary \citep{mckinney02}, but the differences
do not fundamentally modify the solution types.  This scaling of
accretion rate with the location of the inner and outer boundaries is
also roughly consistent with that found by \cite{yuan12a} and
\cite{stone99}.  The transition to the cold disks is also pushed to a
higher Eddington ratio as the inner boundary is moved inwards.  A more
self-consistent approach to determining the exact transition Eddington
ratio would follow infalling gas down to the black hole horizon.

Anpther variable that can affect the properties of the hydrodynamic
accretion flow is the kinematic viscosity.  We have explored other
values of $\nu$ for the hot disk solution at Eddington ratio
$\dot{M}_{\rm B}/\dot{M}_{\rm Edd}=10^{-4}$, and find that the mass
inflow rate from large radii is proportional to $\sqrt{\nu}$.  The
accretion rate as a fraction of the mass inflow rate from large radii
appears to scale linearly with the kinematic viscosity.

\begin{figure}[htp]
\centering
\includegraphics[width=.5\textwidth]{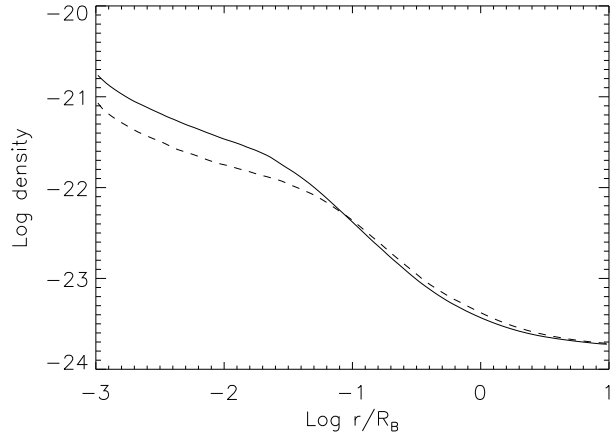}
\caption{Scaling of density with radius for the hot disk solution at
$\dot{M}_{\rm B}/\dot{M}_{\rm Edd}=10^{-4}$.  The solid curve
corresponds to a slice along the equator, and the dashed curve
corresponds to a slice at $\theta=\pi/4$.}\label{fig:densityscaling}
\end{figure}

Finally, in Figure~\ref{fig:densityscaling} we show the scaling of
density with radius for the hot disk solution at $\dot{M}_{\rm
B}/\dot{M}_{\rm Edd}=10^{-4}$.  Densities have been averaged over
times $t=55t_B-90t_B$, and the solid curve represents a slice along
the equator at $\theta=\pi/2$, whereas the dashed curve represents a
slice at constant $\theta=\pi/4$.  The equatorial density scales
roughly as $\rho \propto r^{-0.6}$ in the inner parts of the flow
interior to the centrifugal support radius at $R_c=0.02 R_B$.  Beyond
the centrifugal support radius, $R_c$, the density begins falling more
steeply as $\rho \propto r^{-1.2}$, before levelling off in accordance
with our outer boundary condition.  In comparison, the density scales
as $\rho \propto R^{-3/2}$ in ADAF models \citep{narayan94} and as
$\rho \propto R^{-1/2}$ in standard CDAF models
\citep{quataertgruzinov00,narayan00}, though we use a
different viscosity prescription.

\end{document}